\documentclass[a4paper,manyauthors,nocleardouble,COMPASS]{cernphprep}

\usepackage{graphicx}
\usepackage{subfig}
\usepackage[percent]{overpic}
\usepackage[numbers, square, comma, sort&compress]{natbib}
\usepackage[utf8]{inputenc}

\newcommand{\aphi}{$\phi$\xspace}
\newcommand{\aphis}{$\phi_{s}$\xspace}

\newcommand{\vareps}{$\epsilon$\xspace}
\newcommand{\varEmiss}{$E_{\mathrm{miss}}$\xspace}
\newcommand{\varpttwo}{$p_{T}^{2}$\xspace}
\newcommand{\varQtwo}{$Q^{2}$\xspace}
\newcommand{\varxbj}{$x_{\mathit{Bj}}$\xspace}
\newcommand{\varW}{$W$\xspace}
\newcommand{\varnu}{$\nu$\xspace}
\newcommand{\vart}{$t$\xspace}
\newcommand{\vary}{$y$\xspace}

\newcommand{\assAsfmfs}{$A_{\mathrm{UT}}^{\sin \left( \phi - \phi_{S} \right)}$\xspace}
\newcommand{\assAsdfmfs}{$A_{\mathrm{UT}}^{\sin \left( 2\phi - \phi_{S} \right)}$\xspace}

\newcommand{\assAsfs}{$A_{\mathrm{UT}}^{\sin \phi_{S}}$\xspace}

\newcommand{\rNumIt}[1]{\emph{\romannumeral#1\relax}}

\usepackage{times}
\usepackage{epsfig}
\usepackage{amssymb}
\usepackage{colordvi}
\usepackage{graphicx}
\usepackage{wrapfig,rotating}
\usepackage{amsmath}
\usepackage{hyperref}
\usepackage{caption}
\usepackage{nicefrac}

\usepackage{multicol}
\usepackage{booktabs}
\usepackage{multirow}

\usepackage{tikz}

\usepackage[nodayofweek]{datetime}
\newdateformat{myDateFormat}{\THEDAY\ \monthname[\THEMONTH] \THEYEAR}
\usepackage{lineno}
\begin{document}

\graphicspath{ {pic/} }

\begin{titlepage}
\PHnumber{2016--157}
\PHdate{12 June 2016}
%\PHdate{\myDateFormat \today}

\title{Exclusive $\omega$ meson muoproduction on transversely polarised protons}

\Collaboration{The COMPASS Collaboration}
\ShortAuthor{The COMPASS Collaboration}

\begin{abstract}
Exclusive production of $\omega$ mesons was studied at the COMPASS experiment
by scattering $160~\mathrm{GeV}/\mathit{c}$ muons off transversely polarised
protons. Five single-spin and three double-spin azimuthal asymmetries were
measured in the range of photon virtuality $1~(\mathrm{GeV}/\mathit{c})^2 < Q^2
< 10~(\mathrm{GeV}/\mathit{c})^2$, Bjorken scaling variable $0.003 <
x_{\mathit{Bj}} < 0.3$ and transverse momentum squared of the $\omega$ meson
$0.05~(\mathrm{GeV}/\mathit{c})^2 < p_{T}^{2} <
0.5~(\mathrm{GeV}/\mathit{c})^2$. The measured asymmetries are sensitive to the
nucleon helicity-flip Generalised Parton Distributions (GPD) $E$ that are related to
the orbital angular momentum of quarks, the chiral-odd GPDs $H_{T}$ that are
related to the transversity Parton Distribution Functions, and the sign of the
$\pi\omega$ transition form factor. The results are compared to recent
calculations of a GPD-based model.   
\end{abstract}

%\vfill
%\Submitted{(to be submitted to Nucl. Phys. B)}
\end{titlepage}

{
\pagestyle{empty}
%%%%%%%%%%%%%%%%%%%%%%%%%%%%%%%%%%%%%%%%%%%%%%%%%%%%%%%%%%%%%%%
%
% 2016_auththorlist.tex  
%
%%%%%%%%%%%%%%%%%%%%%%%%%%%%%%%%%%%%%%%%%%%%%%%%%%%%%%%%%%%%%%%
\section*{The COMPASS Collaboration}
\label{app:collab}
\renewcommand\labelenumi{\textsuperscript{\theenumi}~}
\renewcommand\theenumi{\arabic{enumi}}
\begin{flushleft}
C.~Adolph\Irefn{erlangen},
M.~Aghasyan\Irefn{triest_i},
R.~Akhunzyanov\Irefn{dubna}, %phd
M.G.~Alexeev\Irefn{turin_u},
G.D.~Alexeev\Irefn{dubna}, %1
A.~Amoroso\Irefnn{turin_u}{turin_i},
V.~Andrieux\Irefn{saclay},
N.V.~Anfimov\Irefn{dubna}, %2
V.~Anosov\Irefn{dubna}, %3
W.~Augustyniak\Irefn{warsaw}
% ...
\end{flushleft}
%%%%%%%%%%%%%%%%%%%%%%%%%%%%%%%%%%%%%%%%%%%%%%%%%%%%%%%%%%%%%%%%%%%%%%%%%%%%%%%%%%%%%%%%%%%%%%%%%%%%%%%%%%%%%%%%%%%%%%%
%
% institutes
%
%%%%%%%%%%%%%%%%%%%%%%%%%%%%%%%%%%%%%%%%%%%%%%%%%%%%%%%%%%%%%%%%%%%%%%%%%%%%%%%%%%%%%%%%%%%%%%%%%%%%%%%%%%%%%%%%%%%%%%%
%\item \Idef{bielefeld}{Universit\"at Bielefeld, Fakult\"at f\"ur Physik, 33501 Bielefeld, Germany\Arefs{l}}
%\item \Idef{munichlmu}{Ludwig-Maximilians-Universit\"at M\"unchen, Department f\"ur Physik, 80799 Munich, Germany\Arefs{l}\Arefs{r}}
\begin{Authlist}
\item \Idef{turin_p}{University of Eastern Piedmont, 15100 Alessandria, Italy}
\item \Idef{aveiro}{University of Aveiro, Department of Physics, 3810-193 Aveiro, Portugal} 
\item \Idef{bochum}{Universit\"at Bochum, Institut f\"ur Experimentalphysik, 44780 Bochum, Germany\Arefs{l}\Arefs{s}}
\item \Idef{bonniskp}{Universit\"at Bonn, Helmholtz-Institut f\"ur  Strahlen- und Kernphysik, 53115 Bonn, Germany\Arefs{l}}
\item \Idef{bonnpi}{Universit\"at Bonn, Physikalisches Institut, 53115 Bonn, Germany\Arefs{l}}
\item \Idef{brno}{Institute of Scientific Instruments, AS CR, 61264 Brno, Czech Republic\Arefs{m}}
\item \Idef{calcutta}{Matrivani Institute of Experimental Research \& Education, Calcutta-700 030, India\Arefs{n}}
\item \Idef{dubna}{Joint Institute for Nuclear Research, 141980 Dubna, Moscow region, Russia\Arefs{o}}
\item \Idef{erlangen}{Universit\"at Erlangen--N\"urnberg, Physikalisches Institut, 91054 Erlangen, Germany\Arefs{l}}
\item \Idef{freiburg}{Universit\"at Freiburg, Physikalisches Institut, 79104 Freiburg, Germany\Arefs{l}\Arefs{s}}
\item \Idef{cern}{CERN, 1211 Geneva 23, Switzerland}
\item \Idef{liberec}{Technical University in Liberec, 46117 Liberec, Czech Republic\Arefs{m}}
\item \Idef{lisbon}{LIP, 1000-149 Lisbon, Portugal\Arefs{p}}
\item \Idef{mainz}{Universit\"at Mainz, Institut f\"ur Kernphysik, 55099 Mainz, Germany\Arefs{l}}
\item \Idef{miyazaki}{University of Miyazaki, Miyazaki 889-2192, Japan\Arefs{q}}
\item \Idef{moscowlpi}{Lebedev Physical Institute, 119991 Moscow, Russia}
\item \Idef{munichtu}{Technische Universit\"at M\"unchen, Physik Department, 85748 Garching, Germany\Arefs{l}\Arefs{r}}
\item \Idef{nagoya}{Nagoya University, 464 Nagoya, Japan\Arefs{q}}
\item \Idef{praguecu}{Charles University in Prague, Faculty of Mathematics and Physics, 18000 Prague, Czech Republic\Arefs{m}}
\item \Idef{praguectu}{Czech Technical University in Prague, 16636 Prague, Czech Republic\Arefs{m}}
\item \Idef{protvino}{State Scientific Center Institute for High Energy Physics of National Research Center `Kurchatov Institute', 142281 Protvino, Russia}
\item \Idef{saclay}{IRFU, CEA, Université Paris-Saclay, 91191 Gif-sur-Yvette, France\Arefs{s}}
\item \Idef{taipei}{Academia Sinica, Institute of Physics, Taipei 11529, Taiwan}
\item \Idef{telaviv}{Tel Aviv University, School of Physics and Astronomy, 69978 Tel Aviv, Israel\Arefs{t}}
\item \Idef{triest_u}{University of Trieste, Department of Physics, 34127 Trieste, Italy}
\item \Idef{triest_i}{Trieste Section of INFN, 34127 Trieste, Italy}
\item \Idef{triest_ictp}{Abdus Salam ICTP, 34151 Trieste, Italy}
\item \Idef{turin_u}{University of Turin, Department of Physics, 10125 Turin, Italy}
\item \Idef{turin_i}{Torino Section of INFN, 10125 Turin, Italy}
\item \Idef{illinois}{University of Illinois at Urbana-Champaign, Department of Physics, Urbana, IL 61801-3080, USA}   
\item \Idef{warsaw}{National Centre for Nuclear Research, 00-681 Warsaw, Poland\Arefs{u} }
\item \Idef{warsawu}{University of Warsaw, Faculty of Physics, 02-093 Warsaw, Poland\Arefs{u} }
\item \Idef{warsawtu}{Warsaw University of Technology, Institute of Radioelectronics, 00-665 Warsaw, Poland\Arefs{u} }
\item \Idef{yamagata}{Yamagata University, Yamagata 992-8510, Japan\Arefs{q} }
\end{Authlist}
%%%%%%%%%%%%%%%%%%%%%%%%%%%%%%%%%%%%%%%%%%%%%%%%%%%%%%%%%%%%%%%%%%%%%%%%%%%%%%%%%%%%%%%%%%%%%%%%%%%%%%%%%%%%%%%%%%%%%%%
%
% Notes
%
%%%%%%%%%%%%%%%%%%%%%%%%%%%%%%%%%%%%%%%%%%%%%%%%%%%%%%%%%%%%%%%%%%%%%%%%%%%%%%%%%%%%%%%%%%%%%%%%%%%%%%%%%%%%%%%%%%%%%%%
%\item \Adef{a0}{Retired from Universit\"at Bielefeld, Fakult\"at f\"ur Physik, 33501 Bielefeld, Germany}
%\item \Adef{a1}{Present address: Universit\"at Mainz, Helmholtz-Institut f\"ur Strahlen- und Kernphysik, 55099 Mainz, Germany}
%\vspace*{-\baselineskip}
\renewcommand\theenumi{\alph{enumi}}
\begin{Authlist}
\item [{\makebox[2mm][l]{\textsuperscript{*}}}] Deceased
\item \Adef{a}{Also at Instituto Superior T\'ecnico, Universidade de Lisboa, Lisbon, Portugal}
\item \Adef{b}{Also at Department of Physics, Pusan National University, Busan 609-735, Republic of Korea and at Physics Department, Brookhaven National Laboratory, Upton, NY 11973, USA}
\item \Adef{r}{Supported by the DFG cluster of excellence `Origin and Structure of the Universe' (www.universe-cluster.de)}
\item \Adef{d}{Also at Chubu University, Kasugai, Aichi 487-8501, Japan\Arefs{q}}
\item \Adef{x}{Also at Department of Physics, National Central University, 300 Jhongda Road, Jhongli 32001, Taiwan}
\item \Adef{e}{Also at KEK, 1-1 Oho, Tsukuba, Ibaraki 305-0801, Japan}
%\item \Adef{f}{Present address: Universit\"at Bonn, Helmholtz-Institut f\"ur Strahlen- und Kernphysik, 53115 Bonn, Germany}
\item \Adef{g}{Also at Moscow Institute of Physics and Technology, Moscow Region, 141700, Russia}
%\item \Adef{j}{Present address: Typesafe AB, Dag Hammarskj\"olds v\"ag 13, 752 37 Uppsala, Sweden}
\item \Adef{v}{Supported by Presidential grant NSh--999.2014.2}
\item \Adef{h}{Present address: RWTH Aachen University, III.\ Physikalisches Institut, 52056 Aachen, Germany}
\item \Adef{y}{Also at Department of Physics, National Kaohsiung Normal University, Kaohsiung County 824, Taiwan}
\item \Adef{i}{Present address: Uppsala University, Box 516, 75120 Uppsala, Sweden}
\item \Adef{c}{Supported by the DFG Research Training Group Programme 1102  ``Physics at Hadron Accelerators''}
%
% institutes
%
\item \Adef{l}{Supported by the German Bundesministerium f\"ur Bildung und Forschung}
\item \Adef{s}{Supported by EU FP7 (HadronPhysics3, Grant Agreement number 283286)}
\item \Adef{m}{Supported by Czech Republic MEYS Grant LG13031}
\item \Adef{n}{Supported by SAIL (CSR), Govt.\ of India}
\item \Adef{o}{Supported by CERN-RFBR Grant 12-02-91500}
\item \Adef{p}{\raggedright Supported by the Portuguese FCT - Funda\c{c}\~{a}o para a Ci\^{e}ncia e Tecnologia, COMPETE and QREN,
 Grants CERN/FP 109323/2009, 116376/2010, 123600/2011 and CERN/FIS-NUC/0017/2015}
\item \Adef{q}{Supported by the MEXT and the JSPS under the Grants No.18002006, No.20540299 and No.18540281; Daiko Foundation and Yamada Foundation}
\item \Adef{t}{Supported by the Israel Academy of Sciences and Humanities}
\item \Adef{u}{Supported by the Polish NCN Grant 2015/18/M/ST2/00550}
\end{Authlist}

\clearpage
}
\setcounter{page}{1}

%\linenumbers

\section{Introduction}
\label{sec:intro}

Hard exclusive meson production (HEMP) in charged lepton scattering off
nucleons plays an important role in studies of the nucleon structure in terms
of its constituents, \emph{i.e.}\ quarks and gluons. Interest in studying HEMP
as well as deeply virtual Compton scattering (DVCS) has increased recently as
this allows access to generalised parton distributions (GPDs)
\cite{Mueller:1998fv, Ji:1996ek, Ji:1996nm, Radyushkin:1996ru,
Radyushkin:1997ki}, which offer a comprehensive description of the partonic
structure of the nucleon. In particular, GPDs provide a picture of the nucleon
as an extended object \cite{Burkardt:2000za, Burkardt:2002hr, Burkardt:2004bv}.
In this picture, which is often referred to as 3-dimensional nucleon
tomography, longitudinal momenta and transverse spatial degrees of freedom of
partons are correlated. Constraining GPDs may also yield an insight into
angular momenta of quarks, which represent another fundamental property of the
nucleon \cite{Ji:1996ek, Ji:1996nm}. The mapping of nucleon GPDs, which became
one of the key objectives of hadron physics, requires a comprehensive programme
of measuring hard exclusive production of photons and various mesons in a broad
kinematic range. 

The amplitude for hard exclusive meson production by longitudinally polarised
virtual photons was proven to factorise into a hard scattering part that is
calculable in perturbative QCD (pQCD) and a soft part \cite{Radyushkin:1996ru,
Collins:1996fb}. The soft part contains GPDs that describe the structure of the
target nucleon and a distribution amplitude (DA), which accounts for the
structure of the produced meson. The factorisation holds in the limit of large
photon virtuality \varQtwo and large invariant mass \varW of the virtual-photon
nucleon system, but fixed \varxbj, and for $|t|/Q^2\ll1$. Here, \vart is the
squared four-momentum transfer to the proton and
$x_{\mathit{Bj}}=Q^{2}/(2M_{p}\nu)$, where \varnu is the energy of the virtual
photon in the lab frame and $M_p$ is the proton mass. This factorisation is
referred to as `collinear' because parton transverse momenta are neglected. No
similar proof of factorisation exists for transversely polarised virtual
photons. However, phenomenological pQCD-inspired models have been proposed
\cite{Martin:1996bp, Goloskokov:2005sd, Goloskokov:2007nt, Goloskokov:2008ib}
that go beyond the collinear factorisation by postulating the so called
`$k_\perp$ factorisation', where $k_\perp$ denotes the parton transverse
momentum. In the model of Refs.~\cite{Goloskokov:2005sd, Goloskokov:2007nt,
Goloskokov:2008ib, Goloskokov:2013mba, Goloskokov:2014ika}, hereafter referred
to as `GK' model, cross sections and spin-density matrix elements (SDMEs) for
HEMP by both longitudinal and transverse virtual photons can be described
simultaneously.

At leading twist, the chiral-even GPDs $H^f$ and $E^f$, where $f$ denotes a
quark of a given flavour or a gluon, are sufficient to describe exclusive vector
meson production on a spin \nicefrac{1}{2} target. These GPDs are of special
interest as they are related to the total angular momentum carried by partons
in the nucleon \cite{Ji:1996ek}. When higher-twist effects are included in the
DA, the chiral-odd GPDs $H_{T}^{f}$ and $\overline{E}_{T}^{f}$ appear, which
describe the process amplitude with helicity flip of the exchanged quark. They
are also referred to as `transverse' GPDs. While parameterisations of GPDs
$H^{f}$ over the presently accessible \varxbj range are well constrained by
existing measurements of DVCS and HEMP, much less experimental results exist
that allow one to constrain the other mentioned GPDs. For references to
measurements relevant for constraining GPDs $H^{f}$ and $E^{f}$ see \emph{e.g.}\ 
the introductory sections in Refs.~\cite{Adolph:2012ht, Adolph:2013zaa}.
Depending on quark content and quantum numbers of the meson, the soft part of
the process amplitude contains specific combinations of flavour-dependent quark
GPDs and gluon GPDs \cite{Diehl:2003ny, Diehl:2004wj, Goloskokov:2006hr}.
Because of this property HEMP can be regarded as a quark flavour filter, which
motivates the study of a wide spectrum of mesons.

The COMPASS collaboration has already published results on azimuthal
asymmetries for exclusive $\rho^0$ production on transversely polarised protons
\cite{Adolph:2012ht, Adolph:2013zaa} and deuterons \cite{Adolph:2012ht}, which
were compared with predictions of the GPD model of 
Refs.~\cite{Goloskokov:2008ib, Goloskokov:2013mba}. These asymmetries are sensitive
to all types of GPDs, including the chiral-odd GPDs $H_{T}$ and
$\overline{E}_{T}$. In particular, the leading-twist asymmetry \assAsfmfs (see
Sec.~\ref{sec:formalism} for the definition) is sensitive to the chiral-even
GPDs $E$. These GPDs are of special interest, as they describe transitions with
nucleon helicity flip and are related to the orbital angular momentum of
quarks. The model describes well the COMPASS data obtained for $\rho^0$ and
provides their interpretation in terms of GPDs. The measured asymmetry
\assAsfmfs is of small magnitude, because for GPDs $E$ in $\rho^0$ production
the valence quark contribution is expected to be small. This is interpreted as
approximate cancellation due to opposite signs and similar magnitudes of GPDs
$E^{u}$ and $E^{d}$ for valence quarks \cite{Goloskokov:2008ib}. Also, the
small gluon and sea contributions evaluated in Ref.~\cite{Goloskokov:2008ib}
cancel here to a large extent. The model also explains the non-vanishing
asymmetry \assAsfs by a significant contribution from chiral-odd GPDs $H_{T}$
that are related to transversity parton distribution functions. It is the first
experimental indication in hard exclusive $\rho^0$ production of the
contribution of these chiral-odd GPDs.

The interest in studying transverse spin azimuthal asymmetries in hard
exclusive $\omega$ production is twofold. First, due to the different quark
combinations in the flavour-dependent wave functions of the mesons, certain
asymmetries are expected to be larger for $\omega$ production than the
corresponding ones for $\rho^0$. In particular, for \assAsfmfs the version of
the model as described in Ref.~\cite{Goloskokov:2008ib} predicts a sizeable
value of approximately $-0.1$ for the $\omega$ channel in contrast to a small
value predicted for the $\rho^0$ channel. Thus the measurement of this
asymmetry in both channels will provide additional constraints, which may help
to separate the valence quarks contributions $E^{u}$ and $E^{d}$. Secondly, it
is known since a long time that pion exchange can play an important role in
photo- and leptoproduction of $\omega$ mesons \cite{Bauer:1977iq}. The recent
HERMES measurements of SDMEs for exclusive electroproduction of $\omega$ mesons
\cite{Airapetian:2014gfp} indicate a sizeable contribution of the
unnatural-parity-exchange processes in the covered energy range. In the
framework of the GK model it was shown \cite{Goloskokov:2014ika} that the
pion-pole exchange is important to reproduce HERMES results on SDMEs. Still,
SDME data do not allow to distinguish the sign of the $\pi \omega$ transition
form factor. Certain azimuthal asymmetries for $\omega$ production are
sensitive to the pion-pole contribution and hence in principle could allow the
determination of its sign. Although the effect of the pion-pole decreases with
increasing \varW, it might still be measurable beyond experimental
uncertainties at COMPASS. For other vector mesons the effect is expected to be
very small ($\rho^0$ production) or negligible ($\phi$ production)
\cite{Goloskokov:2014ika}. 

This Paper describes the measurement of exclusive $\omega$ muoproduction on
transversely polarised protons with the COMPASS apparatus. Size and kinematic
dependences of azimuthal asymmetries of the cross section with respect to beam
and target polarisation are determined and discussed. The related theoretical
formalism is outlined in the following section. A brief presentation of the
experiment is given in Sec.~\ref{sec:experiment}, while in 
Sec.~\ref{sec:selection} the data selection is reported in detail. The extraction of
asymmetries and the estimation of systematic uncertainties are described in
Sec.~\ref{sec:extraction} and \ref{sec:systematics}, respectively. Results and
concluding remarks are given in Sec.~\ref{sec:results}. 

\section{Theoretical formalism}
\label{sec:formalism}

\begin{figure}[tbp]
\centering
\includegraphics[width=.55\textwidth]{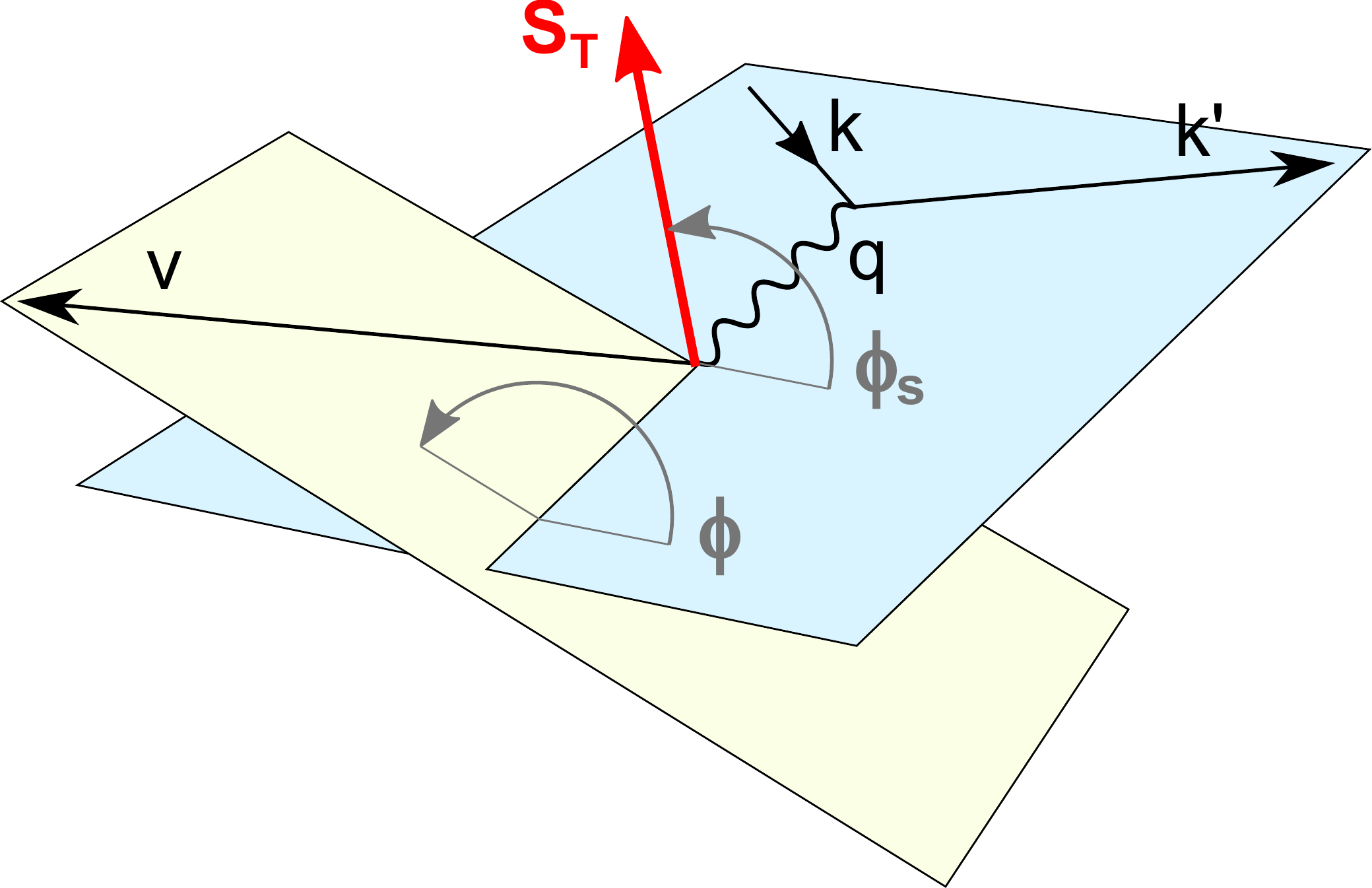}
\caption{Kinematics of exclusive meson production in the target rest frame.
Here $\mathbf{k}$, $\mathbf{k'}$, $\mathbf{q}$ and $\mathbf{v}$ represent the
three-momentum vectors of the incident and the scattered muons, the virtual
photon and the meson respectively. The component of the target spin vector
$\mathbf{S}$ (not shown) perpendicular to the virtual-photon direction is
denoted by $\mathbf{S_T}$.}
\label{fig:formalism:angles}
\end{figure}

The cross section for exclusive $\omega$ muoproduction, $\mu\,N \to
\mu'\,\omega\,N'$, on a transversely polarised nucleon reads
\cite{Diehl:2005pc}:
\footnote{For convenience in this chapter natural units $\hbar = c =1 $ are
used.} 
\footnote{Note that the \vart-dependence of the cross section is indicated explicitly here and the terms $\sigma^{mn}_{ij}$ given by Eq.~\eqref{eq:formalism:sigma} depend on \vart, while in Ref.~\cite{Diehl:2005pc} they are integrated over \vart.} 
\begin{align}
\nonumber 
   & \left[ \frac{\alpha_{\mathit{em}}}{8\pi^{3}} \frac{y^{2}}{1-\epsilon} \frac{1-x_{\mathit{Bj}}}{x_{\mathit{Bj}}} \frac{1}{Q^{2}} \right]^{-1}
\frac{{\mathrm d}\sigma}{{\mathrm d}x_{\mathit{Bj}} {\mathrm d}Q^{2} {\mathrm d}t {\mathrm d}\phi {\mathrm d}\phi_{s}} = &\nonumber \\[8pt]
   &\text{\rule{40pt}{0pt}}    \text{\phantom{${}+{}$}}  \frac{1}{2}\left( \sigma_{++}^{++} + \sigma_{++}^{--} \right) + \epsilon \sigma_{00}^{++}
                               - \epsilon \cos \left( 2\phi \right) \mathrm{Re}\; \sigma_{+-}^{++} &\nonumber \\[2pt] 
   &\text{\rule{40pt}{0pt}}    - \sqrt{\epsilon \left( 1 + \epsilon \right)} \cos \phi\; \mathrm{Re} \left( \sigma_{+0}^{++} + \sigma_{+0}^{--} \right) \text{\phantom{$\Big[$}}&\nonumber \\[2pt]
   &\text{\rule{40pt}{0pt}}-P_{\ell}\sqrt{\epsilon \left( 1 - \epsilon \right)} \sin \phi\; \mathrm{Im} \left( \sigma_{+0}^{++} + \sigma_{+0}^{--} \right) \text{\phantom{$\Big[$}
   }&\nonumber \\[8pt]
   &\text{\rule{40pt}{0pt}}-S_{T}\text{\phantom{$P_{\ell}$}}\Big[
                                 \sin \left( \phi - \phi_{s} \right) \mathrm{Im} \left( \sigma_{++}^{+-} + \epsilon \sigma_{00}^{+-} \right) 
                               + \frac{\epsilon}{2} \sin \left( \phi + \phi_{s} \right) \mathrm{Im}\; \sigma_{+-}^{+-} &\nonumber \\[2pt]
   &\text{\rule{40pt}{0pt}}    \text{\phantom{$S_{T}P_{\ell}$}} + \frac{\epsilon}{2} \sin \left( 3\phi - \phi_{s} \right) \mathrm{Im}\; \sigma_{+-}^{-+} 
                               + \sqrt{\epsilon \left( 1 + \epsilon \right)} \sin \phi_{s}\; \mathrm{Im}\; \sigma_{+0}^{+-} &\nonumber \\[2pt]
   &\text{\rule{40pt}{0pt}}    \text{\phantom{$S_{T}P_{\ell}$}} + \sqrt{\epsilon \left( 1 + \epsilon \right)} \sin \left( 2\phi - \phi_{s} \right) \mathrm{Im}\; \sigma_{+0}^{-+} 
  \Big] &\nonumber \\[8pt]
   &\text{\rule{40pt}{0pt}}+S_{T} P_{\ell} \Big[
                                 \sqrt{ 1 - \epsilon^{2}} \cos \left( \phi - \phi_{s} \right) \mathrm{Re}\; \sigma_{++}^{+-}
                               - \sqrt{\epsilon \left( 1 - \epsilon \right)} \cos \phi_{s}\; \mathrm{Re}\; \sigma_{+0}^{+-} &\nonumber \\[2pt]
   &\text{\rule{40pt}{0pt}}    \text{\phantom{$S_{T}P_{\ell}$}} - \sqrt{\epsilon \left( 1 - \epsilon \right)} \cos \left( 2\phi - \phi_{s} \right) \mathrm{Re}\; \sigma_{+0}^{-+} 
   \Big] , &
\label{eq:formalism:cross_section}
\end{align}
where only terms relevant for the present analysis are shown. For brevity, the
dependence on kinematic variables is omitted. The general formula for the cross
section for meson leptoproduction can be found in Ref.~\cite{Diehl:2005pc}. The
angle \aphi is the azimuthal angle between the lepton plane that is spanned by
the momenta of the incoming and the scattered leptons, and the hadron plane
that is spanned by the momenta of the virtual photon and the meson (see 
Fig.~\ref{fig:formalism:angles}). The angle \aphis is the azimuthal angle between
the lepton plane and the spin direction of the target nucleon.

The polarisation of the lepton beam is denoted by $P_{\ell}$. The component of
the transverse target spin perpendicular to the virtual-photon direction,
$S_{T}$, is approximated in the COMPASS kinematic region by the corresponding
component perpendicular to the direction of the incoming muon, $P_{T}$.
According to Ref.~\cite{Diehl:2005pc}, the transition from $S_{T}$ to $P_{T}$
introduces in Eq.~\eqref{eq:formalism:cross_section} a dependence on $\theta$,
which is the angle between the directions of virtual photon and incoming beam
particle. This dependence gives rise to additional asymmetries of the cross
section that are related to longitudinal target polarisation. These asymmetries
are suppressed by the factor $\sin \theta$, which is small at COMPASS
kinematics ($\sin\theta \approx 0.056$ on average). In the present analysis the effect of the angle $\theta$ is
neglected.

In the considered kinematics, where the mass of the incoming lepton $m_{\mu}
\ll Q^{2}$, the virtual-photon polarisation parameter \vareps can be
approximated in the following way:
\begin{equation}
\epsilon \approx \frac{\displaystyle 1-y-\frac{1}{4}y^2\gamma ^2}{\displaystyle 1-y+\frac{1}{2}y^2+\frac{1}{4}y^2\gamma ^2}. 
\label{eq:formalism:epsilon}
\end{equation}
Here, \vary is the fractional energy of the virtual photon (see 
Table~\ref{tab:formalism:kinetic_variables}), $\gamma^2 = \left( 2 x_{\mathit{Bj}}
M_{p} \right)^2/Q^{2}$ and $M_p$ is the mass of the proton.

The photoabsorption cross sections or interference terms $\sigma_{ij}^{mn}$ are
proportional to bilinear combinations of helicity amplitudes $\mathcal{M}$ for
the photoproduction subprocess,
\begin{equation}
\sigma_{ij}^{mn} \propto \sum \mathcal{M}_{i'm',im}^{*}\mathcal{M}_{i'm',jn},
\label{eq:formalism:sigma}
\end{equation}
where the helicity of the virtual photon is denoted by $i, j = -1, 0, +1$ and
the helicity of the initial-state proton is denoted by $m, n =
-\nicefrac{1}{2}, +\nicefrac{1}{2}$. The sum runs over all combinations of
helicities of meson ($i' = -1, 0, +1$) and final-state proton ($m' =
-\nicefrac{1}{2}, +\nicefrac{1}{2}$). In the following the helicities $-1,
-\nicefrac{1}{2}, 0, +\nicefrac{1}{2}, +1$ will be labelled by only their sign
or zero, omitting $\nicefrac{1}{2}$ or $1$.

For a transversely polarised target five single ($\mathrm{UT}$) and three
double ($\mathrm{LT}$) spin asymmetries can be defined: 
\begin{align}
 A_{\mathrm{UT}}^{\sin \left( \phi - \phi_{s} \right)\text{\phantom{$3$}}} &= - \frac{\mathrm{Im} \left( \sigma_{++}^{+-} + \epsilon \sigma_{00}^{+-} \right)}{\sigma_{0}}, 
&A_{\mathrm{LT}}^{\cos \left( \phi - \phi_{s} \right)\text{\phantom{$3$}}} &= \phantom{-}  \frac{\mathrm{Re}\; \sigma_{++}^{+-}}{\sigma_{0}}, \nonumber\\
 A_{\mathrm{UT}}^{\sin \left( \phi + \phi_{s} \right)\text{\phantom{$3$}}} &= - \frac{\mathrm{Im}\; \sigma_{+-}^{+-}}{\sigma_{0}},
&A_{\mathrm{LT}}^{\cos \phi_{s}\text{\phantom{$\left( 3\phi + \right) $}}} &= - \frac{\mathrm{Re}\; \sigma_{+0}^{+-}}{\sigma_{0}}, \nonumber\\
 A_{\mathrm{UT}}^{\sin \left(3\phi - \phi_{s} \right)} &= - \frac{\mathrm{Im}\; \sigma_{+-}^{-+}}{\sigma_{0}},
&A_{\mathrm{LT}}^{\cos \left(2\phi - \phi_{s} \right)} &= - \frac{\mathrm{Re}\; \sigma_{+0}^{-+}}{\sigma_{0}}, \nonumber\\
 A_{\mathrm{UT}}^{\sin \phi_{s}\text{\phantom{$\left( 3\phi + \right) $}}} &= - \frac{\mathrm{Im}\; \sigma_{+0}^{+-}}{\sigma_{0}}, \nonumber\\
 A_{\mathrm{UT}}^{\sin \left(2\phi - \phi_{s} \right)} &= - \frac{\mathrm{Im}\; \sigma_{+0}^{-+}}{\sigma_{0}}.        
\label{eq:formalism:asymmetries}
\end{align}
Here, $\sigma_{0}$ is the total unpolarised cross section, which is the sum of
the cross sections for longitudinally and transversely polarised virtual
photons, $\sigma_{L}$ and $\sigma_{T}$, respectively: 
\begin{equation}
\sigma_{0} = \frac{1}{2}\left( \sigma_{++}^{++} + \sigma_{++}^{--} \right) + \epsilon \sigma_{00}^{++} = \sigma_{T} + \epsilon \sigma_{L} . 
\label{eq:formalism:sigma0}
\end{equation}
Each asymmetry is related to a modulation of the cross section as a function of
\aphi and/or \aphis (see Eq.~\eqref{eq:formalism:cross_section}), which is
indicated by the superscript. 

Calculations for the full set of five $A_{\mathrm{UT}}$ and three
$A_{\mathrm{LT}}$ asymmetries were performed recently in the framework of the
GK model \cite{Goloskokov:2013mba}. Of particular interest for an
interpretation of the COMPASS results described in this Paper are three
asymmetries, which can be expressed through helicity amplitudes neglecting
terms containing unnatural parity exchange amplitudes:
\begin{align}
\sigma_{0}\; A_{\mathrm{UT}}^{\sin \left( \phi - \phi_{s} \right)\text{\phantom{$3$}}}& =
-2 \mathrm{Im} \left[ \epsilon \mathcal{M}_{0-,0+}^{*} \mathcal{M}_{0+,0+} + \mathcal{M}_{+-,++}^{*} \mathcal{M}_{++,++} + \tfrac{1}{2} \mathcal{M}_{0-,++}^{*} \mathcal{M}_{0+,++} \right]~, \nonumber \\[0pt]
\sigma_{0}\; A_{\mathrm{UT}}^{\sin \left(2\phi - \phi_{s} \right)} & =
-\phantom{2} \mathrm{Im} \left[ \mathcal{M}_{0+,++}^{*} \mathcal{M}_{0-,0+} \right]~, \nonumber \\[0pt]
\sigma_{0}\; A_{\mathrm{UT}}^{\sin \phi_{s}\text{\phantom{$\left( 3\phi + \right) $}}}& =
-\phantom{2} \mathrm{Im} \left[ \mathcal{M}_{0-,++}^{*} \mathcal{M}_{0+,0+} - \mathcal{M}_{0+,++}^{*} \mathcal{M}_{0-,0+} \right]~.
\end{align}
Most of the neglected amplitudes are related to pion pole exchange, the role of
which will be discussed in Sec.~\ref{sec:results}. 

The dominant contribution from the $\gamma^{*}_{L} \rightarrow V_{L}$
transition, where $V$ denotes vector meson, is described by
$\mathcal{M}_{0+,0+}$ and $\mathcal{M}_{0-,0+}$, which are related to
chiral-even GPDs $H$ and $E$. The suppressed contribution from the
$\gamma^{*}_{T} \rightarrow V_{T}$ transition is described by
$\mathcal{M}_{++,++}$ and $\mathcal{M}_{+-,++}$, which are also related to
chiral-even GPDs. A description of the $\gamma^{*}_{T} \rightarrow V_{L}$
transition is possible by including chiral-odd GPDs $H_{T}$ and $\overline
E_{T}$, which are related to $\mathcal{M}_{0-,++}$ and $\mathcal{M}_{0+,++}$,
respectively. The $\gamma^{*}_{L} \rightarrow V_{T}$ and $\gamma^{*}_{T}
\rightarrow V_{-T}$ transitions are known to be suppressed and are neglected
here.

Different values are predicted for the asymmetry \assAsfmfs in $\rho^0$ and
$\omega$ productions, as already mentioned above. For this asymmetry, the
contribution of chiral-odd GPDs is expected to be negligible, as one can see
for instance from the comparison of calculations for the $\rho^0$ channel in
Refs.~\cite{Goloskokov:2008ib} and \cite{Goloskokov:2013mba}. The asymmetry
\assAsfs represents an imaginary part of two bilinear products of helicity
amplitudes. The first product is related to GPDs $H$ and $H_{T}$, while the
second one is related to GPDs $E$ and $\overline E_{T}$. The latter product
appears also in the asymmetry \assAsdfmfs. For the $\rho^0$ channel the
asymmetry \assAsfs was found to be different from zero, while the asymmetry
\assAsdfmfs is compatible with zero \cite{Adolph:2013zaa}. This implies a
non-negligible contribution of GPDs $H_{T}$ in this case.

A summary of the kinematic variables used in this Paper is given in 
Table~\ref{tab:formalism:kinetic_variables}. 

\begin{table}[t]
\centering
\caption{Kinematic variables.}
\label{tab:formalism:kinetic_variables}
{\small
\begin{tabular}{l l}
\toprule
$k$                                                &  four-momentum of incident muon \\
$k'$                                               &  four-momentum of scattered muon \\
$p$                                                &  four-momentum of target nucleon \\
$v$                                                &  four-momentum of $\omega$ meson \\
$q=k-k'$                                           &  four-momentum of virtual photon \\
$Q^2=-q^2$                                         &  negative invariant mass squared of virtual photon \\
$W=\sqrt{(p+q)^2}$                                 &  invariant mass of the $\gamma^{*}-N$ system \\
$M_p$                                              &  proton mass \\
$\nu=(p \cdot q)/M_p$                              &  energy of virtual photon in the laboratory system \\
$x_{\mathit{Bj}}=Q^2/(2M_p \nu)$                   &  Bjorken scaling variable \\
$y=(p \cdot q)/(p \cdot k)$                        &  fraction of lepton energy lost in the laboratory system \\
$M_{\pi\pi\pi}$                                    &  invariant mass of $\pi^+\pi^-\pi^0$ system \\
$t=(q-v)^2$                                        &  square of the four-momentum transfer to the target nucleon \\
$p_T^2$                                            &  transverse momentum squared of vector meson with \\
                                                   &  respect to the virtual-photon direction \\
$E_{\omega}$                                       &  energy of $\omega$ in the laboratory system \\
$M_X^2=(p+q-v)^2$                                  &  missing mass squared of the undetected system \\
$E_{\textrm{miss}}=((p+q-v)^2 - p^2)/(2M_p)$       &  missing energy of the undetected system \\
$\phantom{E_{\textrm{miss}}}=({M_X^2 - M_p^2})/({2M_p})$
&  \\
$\phantom{E_{\textrm{miss}}}=\nu - E_{\omega} + t/(2 M_p)$
&  \\
\bottomrule
\end{tabular}
}
\end{table}
 
\section{Experimental set-up}
\label{sec:experiment}

COMPASS is a fixed-target experiment situated at the high-intensity M2 beam
line of the CERN SPS. A detailed description of the experiment can be found in
Ref.~\cite{Abbon:2007pq}.

The $\mu^+$ beam had a nominal momentum of $160~\mathrm{GeV}/\mathit{c}$ with a
spread of $5\%$ and a longitudinal polarisation of $P_{\ell}\approx-0.8$ known
with the precision of $5\%$. The data were taken at a mean intensity of $3.5
\times 10^{8}~\mu/\mathrm{spill}$, for a spill length of about $10~\mathrm{s}$
every $40~\mathrm{s}$. A measurement of the trajectory and the momentum of each
incoming muon is performed upstream of the target. The momentum of the beam
muon is measured with a relative precision better than $1\%$. 

The beam traverses a solid-state ammonia ($\mathrm{NH}_3$) target that contains
transversely polarised protons. The target is situated within a large aperture
magnet with a dipole holding field of $0.5~\mathrm{T}$. The $2.5~\mathrm{T}$
solenoidal field is only used when polarising the target material. A mixture of
liquid $^3\mathrm{He}$ and $^4\mathrm{He}$ is used to cool the target to
$50~\mathrm{mK}$. Ten nuclear magnetic resonance (NMR) coils surrounding the
target allow for a determination of the target polarisation $P_{T}$, which
typically amounts to $0.8$ with an uncertainty of $3\%$. The ammonia is
contained in three cylindrical target cells with a diameter of $4~\mathrm{cm}$,
placed along the beam with $5~\mathrm{cm}$ space between cells. The central
cell is $60~\mathrm{cm}$ long and the two outer ones are $30~\mathrm{cm}$ long.
The spin directions in neighbouring cells are opposite. Such a target
configuration allows for a simultaneous measurement of azimuthal asymmetries
for the two target spin directions without relying on beam flux measurements.
Systematic effects due to acceptance are reduced by reversing the spin
directions on a weekly basis. With the three-cell configuration, the average
acceptance for cells with opposite spin direction is approximately the same,
which leads to a further reduction of systematic uncertainties.

The dilution factor $f$, which is the cross-section-weighted fraction of
polarisable material, is calculated for incoherent exclusive $\omega$
production using the measured material composition and the nuclear dependence
of the cross section:
\begin{equation}
f = \frac{n_p}{n_p + \sum_A {n_A \frac{\widetilde{\sigma}_A}{\sigma_p}}} .
\label{eq:experiment:dilution_factor}
\end{equation}
Here, $n_p$ and $n_A$ denote the numbers of polarisable protons in the target
and of unpolarised nucleons in the target material with atomic mass $A$,
respectively. The sum runs over all nuclei present in the COMPASS target. The
ratio of the cross section per nucleon for a given nucleus to the cross section
on the proton is denoted by $\widetilde{\sigma}_{A} / \sigma_{p}$. The
inclusion of the effect of nuclear shadowing on the calculation of the dilution
factor is crucial for the ammonia target. However, this effect has never been
measured for exclusive $\omega$ production in a kinematic region comparable to
that covered by the COMPASS experiment. Therefore, we assume that the nuclear
shadowing effect for $\omega$ is the same as that for $\rho^0$. This assumption
is supported by similar quark compositions, quantum numbers ($J^P$) and masses
of both mesons. The assumption leads to the same dilution factor as for
$\rho^0$, the evaluation of which is detailed in Ref.~\cite{Alexakhin:2007mw}.
For the $\mathrm{NH}_3$ target, which is used for the present analysis, the
dilution factor amounts typically to $0.25$ \cite{Adolph:2012ht}.

The COMPASS spectrometer is designed to reconstruct scattered muons and
produced hadrons in wide momentum and angular ranges. It consists of two
stages, each equipped with a dipole magnet, to measure tracks with large and
small momenta, respectively.  In the high-flux region, in or close to the beam,
tracking is provided by stations of scintillating fibres, silicon detectors,
micromesh gaseous chambers and gas electron multiplier chambers. Large-angle
tracking devices are multiwire proportional chambers, drift chambers and straw
detectors. Muons are identified in large-area mini drift tubes and drift tubes
placed downstream of hadron absorbers. Each stage of the spectrometer contains
an electromagnetic and a hadron calorimeter. The identification of charged
particles is possible with a RICH detector, although in this analysis it is not
used.

The data recording system is activated by several triggers. For inclusive
triggers, the scattered muon is identified by a coincidence of signals from
trigger hodoscopes. Semi-inclusive triggers select events with a scattered muon
and an energy deposit in a hadron calorimeter exceeding a given threshold.
Moreover, a pure calorimeter trigger with a high energy threshold was
implemented to extend the acceptance towards high \varQtwo and large \varxbj.
It was checked that this trigger does not introduce any bias due to the
acceptance of the calorimeters in the \varxbj range covered by the present
data. Veto counters upstream of the target are used to suppress beam halo
muons.

\section{Event sample}
\label{sec:selection}

The results presented in this Paper are based on the data taken with the
transversely polarised $\mathrm{NH}_3$ target in 2010. An event to be accepted
for further analysis is required to have the same topology as that of the
observed process
\begin{center}
\begin{tikzpicture}
\node [right] at (0., 1.2) {$\mu N \rightarrow \mu' N' \omega$};
\node [right] at (3.0, 0.6) {$\pi^+ \pi^- \pi^0$};
\node [right] at (5.1, 0.) {$\gamma \gamma$~~.};
\draw [thick, ->] (2.1, 0.6+0.3) -- (2.1, 0.6) -- (2.1+0.6, 0.6);
\draw [thick, ->] (4.2, 0.3) -- (4.2, 0.) -- (4.2+0.6, 0.);
\node [right] at (7.2, 0.3) {$\mathrm{BR} \approx 99\%$};
\node [right] at (7.2, 0.6+0.3) {$\mathrm{BR} \approx 89\%$};
\label{eq:selection:reaction}
\end{tikzpicture}
\end{center}
Therefore, we select only events that have an incident muon track, a scattered
muon track, exactly two additional tracks of oppositely charged hadrons, which
are all associated to a vertex in the polarised target material, and a single
$\pi^0$ meson that is reconstructed using its two decay photons detected in the
electromagnetic calorimeters. 

\begin{figure}[tbp]
\centering
\includegraphics[width=.55\textwidth,trim=0 0 0 15]{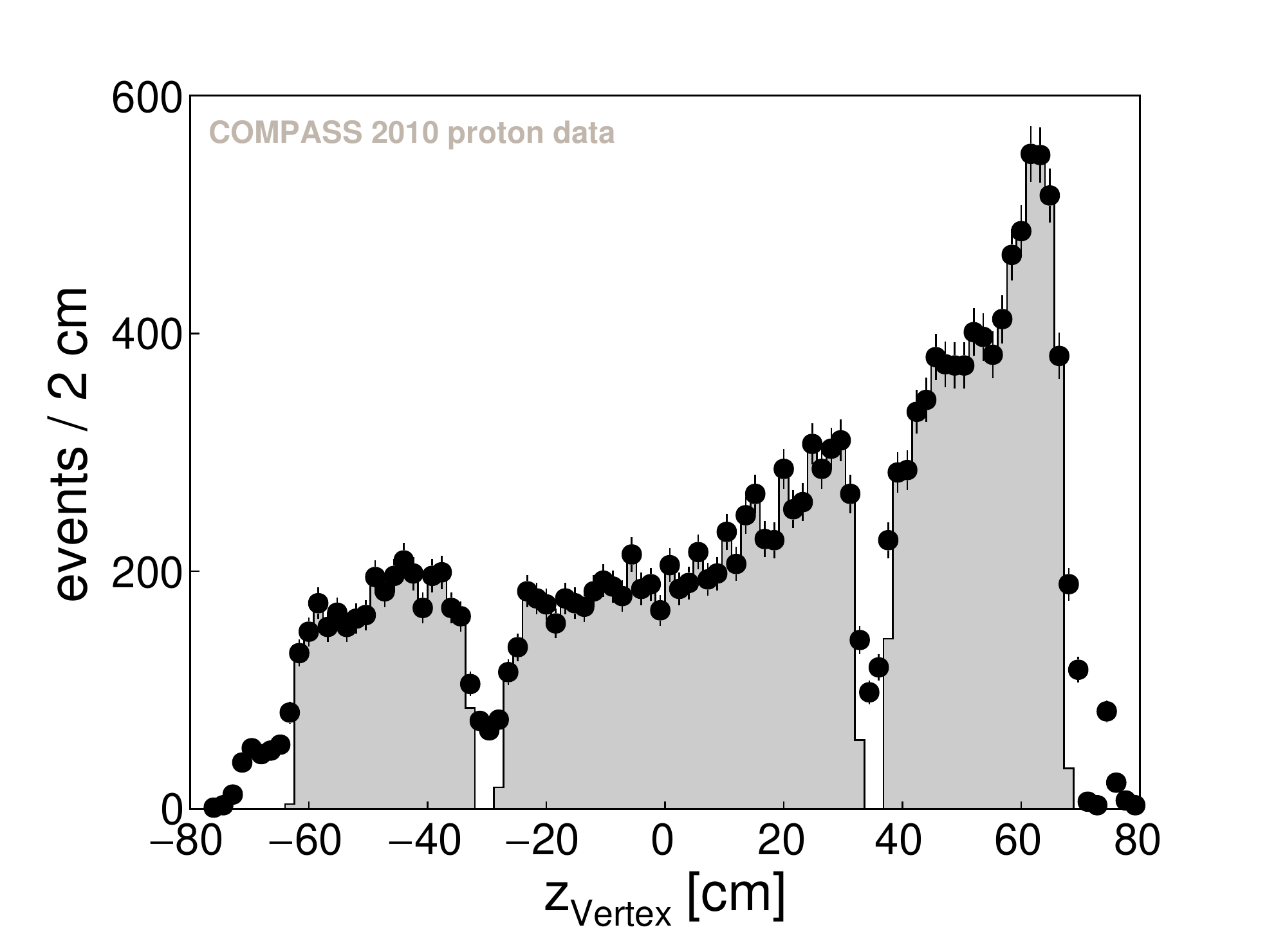}
\caption{Distribution of z-coordinate of reconstructed  primary vertices. The accepted events are denoted by the shaded area.}
\label{fig:selection:zvtx}
\end{figure}

The flux of the incoming beam is equalised for all target cells using
appropriate cuts on position and angle of beam tracks. 
Figure~\ref{fig:selection:zvtx} shows the distribution of the reconstructed vertex
position $z_{V}$ along the beam axis. In this figure as well as in 
Figs.~\ref{fig:selection:gamma_gamma_mass} to \ref{fig:extraction:emiss_fits}, the
distributions are obtained applying all selections except that corresponding to
the displayed variable.

In order to obtain a data sample in the deep inelastic scattering region, the
following kinematic cuts are applied: $1~(\mathrm{GeV}/\mathit{c})^2 < Q^2 <
10~(\mathrm{GeV}/\mathit{c})^2$, where the lower cut selects the perturbative
QCD region and the upper one is chosen to remove the region of \varQtwo where
the fraction of non-exclusive background is large; $0.1 < y < 0.9$, in order to
suppress radiative corrections (large \vary) or poorly reconstructed kinematics
(low \vary). The latter cut removes also events from the region of hadron
resonances at small values of \varW.  A small residual number of such events is
removed by requiring \varW to be larger than $5~\mathrm{GeV}/\mathit{c}^2$.

\subsection{Reconstruction of $\pi^0$}
\label{sec:selection:pi0}

A neutral pion is reconstructed using clusters in the electromagnetic calorimeters (ECALs), which are required not to be associated to charged particle tracks. Only events with two such clusters, which have to pass the selections described below, are retained for the analysis. The
possibility to reconstruct $\pi^0$ mesons by using events with more than two
clusters and examining all cluster combinations was checked in 
Ref.~\cite{Wolbeek:2015}. As such combinatorial method would lead to an increase of
background by more than a factor of two, it is not applied in this analysis.

A photon reconstructed in a given ECAL is accepted only if its energy $E_{\gamma}$ is in the range
\begin{align}
0.6~\mathrm{GeV} < E_{\gamma} < 25 ~\mathrm{GeV} ~~~ \mathrm{for~ ECAL1}, \nonumber \\
1.0~\mathrm{GeV} < E_{\gamma} < 50 ~\mathrm{GeV} ~~~ \mathrm{for~ ECAL2}.
\label{eq:selection:gamma_energy_limits}
\end{align}
Here, ECAL1 (ECAL2) denotes the electromagnetic calorimeter in the large
(small) angle stage of the spectrometer. The yields of exclusive $\omega$
mesons were studied as a function of the values of the lower limits on
$E_{\gamma}$ resulting in maximal yields for the indicated values. The purity
of the exclusive $\omega$ sample only weakly depends on these lower limits. The
upper limits on $E_{\gamma}$ are determined by requiring sufficient statistics
needed for a reliable determination of the $E_{\gamma}$-dependent
parameterisation of the time correlation between a given decay photon candidate
and the incoming muon track. In order to ensure this correlation, the difference
of the measured ECAL cluster time and the measured time of the incoming muon,
$\Delta t = t_{\gamma} - t_{\mu}$, is calculated. Since the precision of time
reconstruction in ECALs depends on the cluster energy, the time correlation is
ensured by requiring 
\begin{equation}  
|\Delta t - \Delta t_{\mathrm{par}}(E_{\gamma})| < 3~ \sigma_{\mathrm{par}}(E_{\gamma})~. 
\end{equation}
For each calorimeter, position $\Delta t_{\mathrm{par}}(E_{\gamma})$ and width $\sigma_{\mathrm{par}}(E_{\gamma})$ of the $\gamma-\mu$ correlation peak are parameterised as a function of $E_{\gamma}$ using a sample of events for semi-inclusive $\pi^0$ production. 

\begin{figure}[tbp]
\centering
\includegraphics[width=.55\textwidth,trim=0 0 0 15]{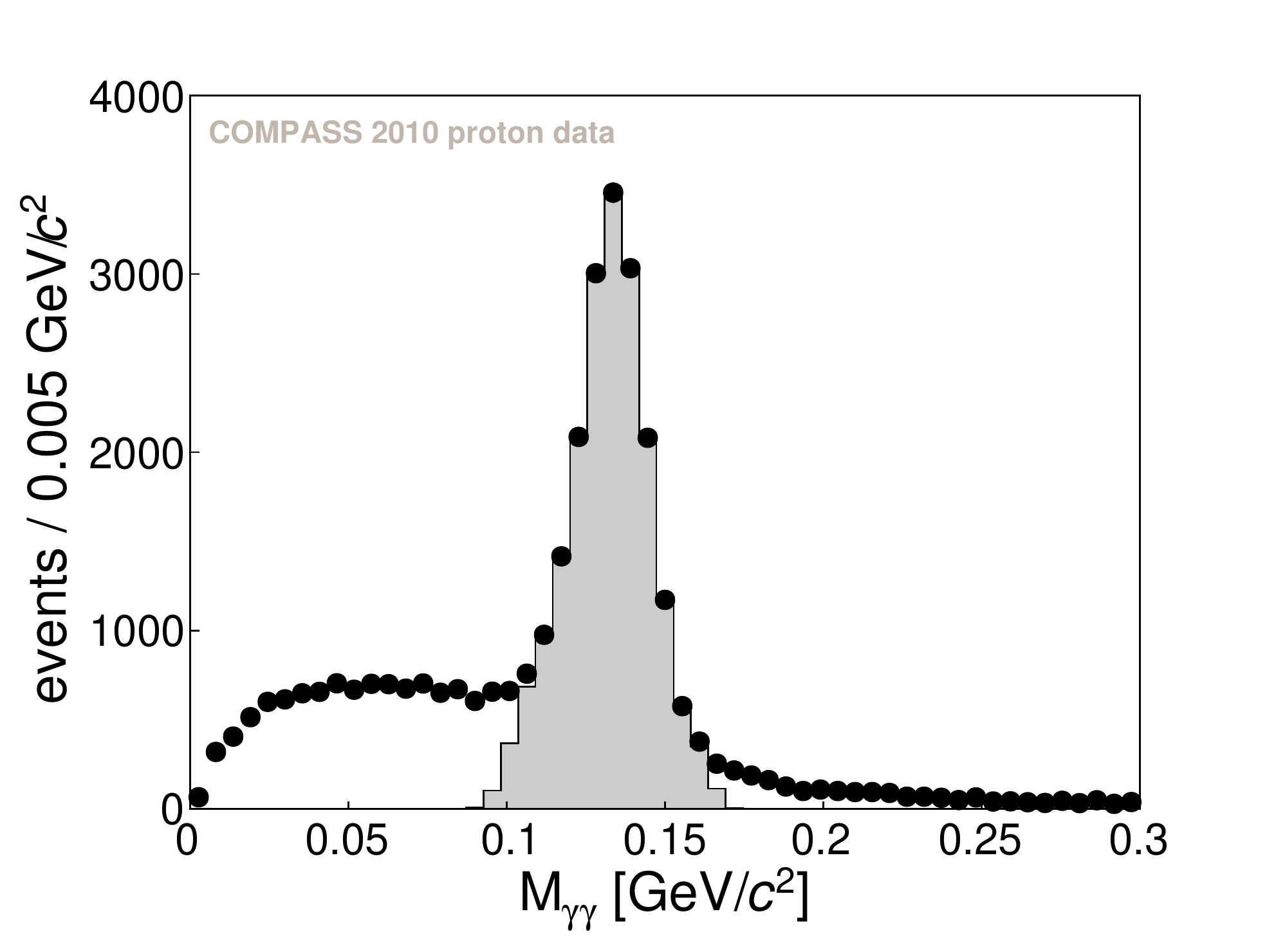}
\caption{Distribution of the invariant mass of two photons. The accepted events are denoted by the shaded area.}
\label{fig:selection:gamma_gamma_mass}
\end{figure}

Similarly, the limit on the invariant mass of two photons, $M_{\gamma\gamma}$,
depends on the energy $E_{\gamma\gamma}$ of the $\pi^0$ candidate:
\begin{equation}  
|M_{\gamma\gamma} - M_{\pi^0,\; \mathrm{par}}(E_{\gamma\gamma})| < 3~ \sigma_{\mathrm{par}}(E_{\gamma\gamma})~. 
\label{eq:selection:pi0_selection_vs_energy}
\end{equation}
Also here, position $M_{\pi^0,\; \mathrm{par}}(E_{\gamma\gamma})$ and width
$\sigma_{\mathrm{par}}(E_{\gamma\gamma})$ of the $\pi^0$ peak are parameterised
using semi-inclusive data for $\pi^0$ mesons reconstructed in each of the three
possible combinations of neutral clusters in ECALs. In addition to the real
data, similar parameterisations are obtained also for Monte Carlo data that are
used for the procedure of background subtraction, see 
Sec.~\ref{sec:extraction}. The parameterisations are obtained in the following
ranges of energy:
\begin{align}
1.2 ~\mathrm{GeV} &< E_{\gamma\gamma} < 25 ~\mathrm{GeV} ~~~ \mathrm{for~ ECAL1}, \nonumber \\ 
2.0 ~\mathrm{GeV} &< E_{\gamma\gamma} < 50 ~\mathrm{GeV} ~~~ \mathrm{for~ ECAL2}, \nonumber \\ 
1.6 ~\mathrm{GeV} &< E_{\gamma\gamma} < 35 ~\mathrm{GeV} ~~~ \mathrm{for~ ECAL1+ECAL2}. 
\label{eq:selection:gamma_gamma_energy_limits}
\end{align}
The selection of $\pi^0$ mesons is restricted to the ranges of energy given in
Eq.~\eqref{eq:selection:gamma_gamma_energy_limits}. The distribution of
$M_{\gamma\gamma}$ for reconstructed events is shown in 
Fig.~\ref{fig:selection:gamma_gamma_mass}, where the accepted events are represented
by the shaded histogram. Note that there are no sharp limits on this histogram,
because the energy-dependent selection on $M_{\gamma\gamma}$ is applied, see
Eq.~\eqref{eq:selection:pi0_selection_vs_energy}. 

In order to reduce the smearing related to ECAL reconstruction, after having
performed the $\pi^0$ selection the energies of decay photons for each event
are rescaled by the factor
\begin{equation}
f_{E_{\gamma}} = \frac{M_{\pi^0}^{\mathrm{PDG}}}{M_{\gamma\gamma}},
\end{equation}
where $M_{\pi^0}^{\mathrm{PDG}} \approx 0.135~\mathrm{GeV}/\mathit{c}^{2}$ is
the nominal $\pi^0$ mass. This reduces the width $\sigma_{\omega}$ of the reconstructed $\omega$
resonance from $25~\mathrm{MeV}/\mathit{c}^{2}$ to
$20~\mathrm{MeV}/\mathit{c}^{2}$.

\subsection{Selection of incoherent exclusive $\omega$ production}
\label{sec:selection:omega}

Events corresponding to incoherent exclusive $\omega$ production are selected
using additional cuts on:
\begin{itemize}
\item{
   the invariant mass of the $\pi^+\pi^-\pi^0$ system, $M_{\pi^+\pi^-\pi^0}$,
   \begin{equation}
   \left| M_{\pi^+\pi^-\pi^0} - M_{\omega}^{\mathrm{PDG}} \right| < 60~\mathrm{MeV}/\mathit{c}^2 , 
   \label{eq:selection:3pi_limit}
   \end{equation}
   where $M_{\omega}^{\mathrm{PDG}} = 782.65~\mathrm{MeV}/\mathit{c}^2$ is the
nominal $\omega$ resonance mass;\\
}
\item{
   the missing energy \varEmiss, 
   \begin{equation}
   -3.0~\mathrm{GeV} < E_{\textrm{miss}} < 3.0~\mathrm{GeV}; 
   \label{eq:selection:Emiss_limit}
   \end{equation}
}
\item{
   the $\omega$ meson energy in the laboratory system, 
   \begin{equation}
   E_{\omega} > 14~\mathrm{GeV}; 
   \label{eq:selection:Eomega_limit}
   \end{equation}
}
\item{
   the transverse momentum squared of the $\omega$ meson with respect to the virtual-photon direction,
   \begin{equation}
   0.05~(\mathrm{GeV}/\mathit{c})^2 < p_{T}^{2} < 0.5~(\mathrm{GeV}/\mathit{c})^2 .
   \label{eq:selection:pt2_limit}
   \end{equation}
}
\end{itemize}

\begin{figure}[tbp]
\centering
\includegraphics[width=.55\textwidth,trim=0 0 0 15]{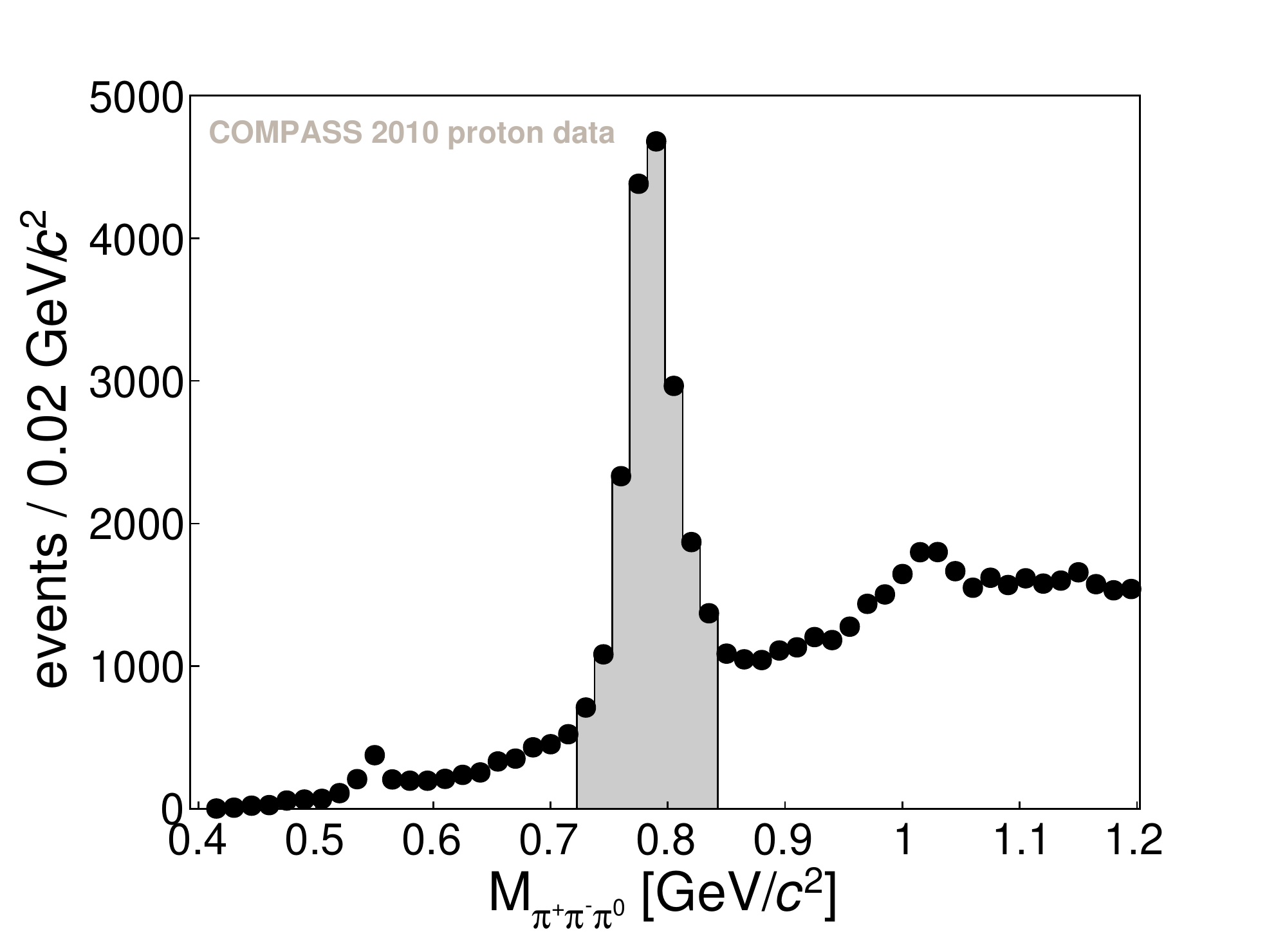}
\caption{Distribution of $M_{\pi^{+}\pi^{-}\pi^{0}}$. The accepted events are denoted by the shaded area.}
\label{fig:selection:3pi_mass}
\end{figure}

The $\omega$ meson is reconstructed using two charged hadrons and a
reconstructed $\pi^0$. As RICH information is not used in this analysis, the
charged pion mass hypothesis is assigned to each hadron track. 
Figure~\ref{fig:selection:3pi_mass} shows the corresponding invariant mass spectrum
that indicates clearly the $\omega$ signal at the nominal position,
$M_{\omega}^{\mathrm{PDG}} = 782.65~\mathrm{MeV}/\mathit{c}^2$. The selection
of $\omega$ mesons using the invariant mass range given by
Eq.~\eqref{eq:selection:3pi_limit} corresponds to the $\pm 3\sigma_{\omega}$ region around
$M_{\omega}^{\mathrm{PDG}}$.

\begin{figure}[tbp]
\centering
\includegraphics[width=.55\textwidth,trim=0 0 0 15]{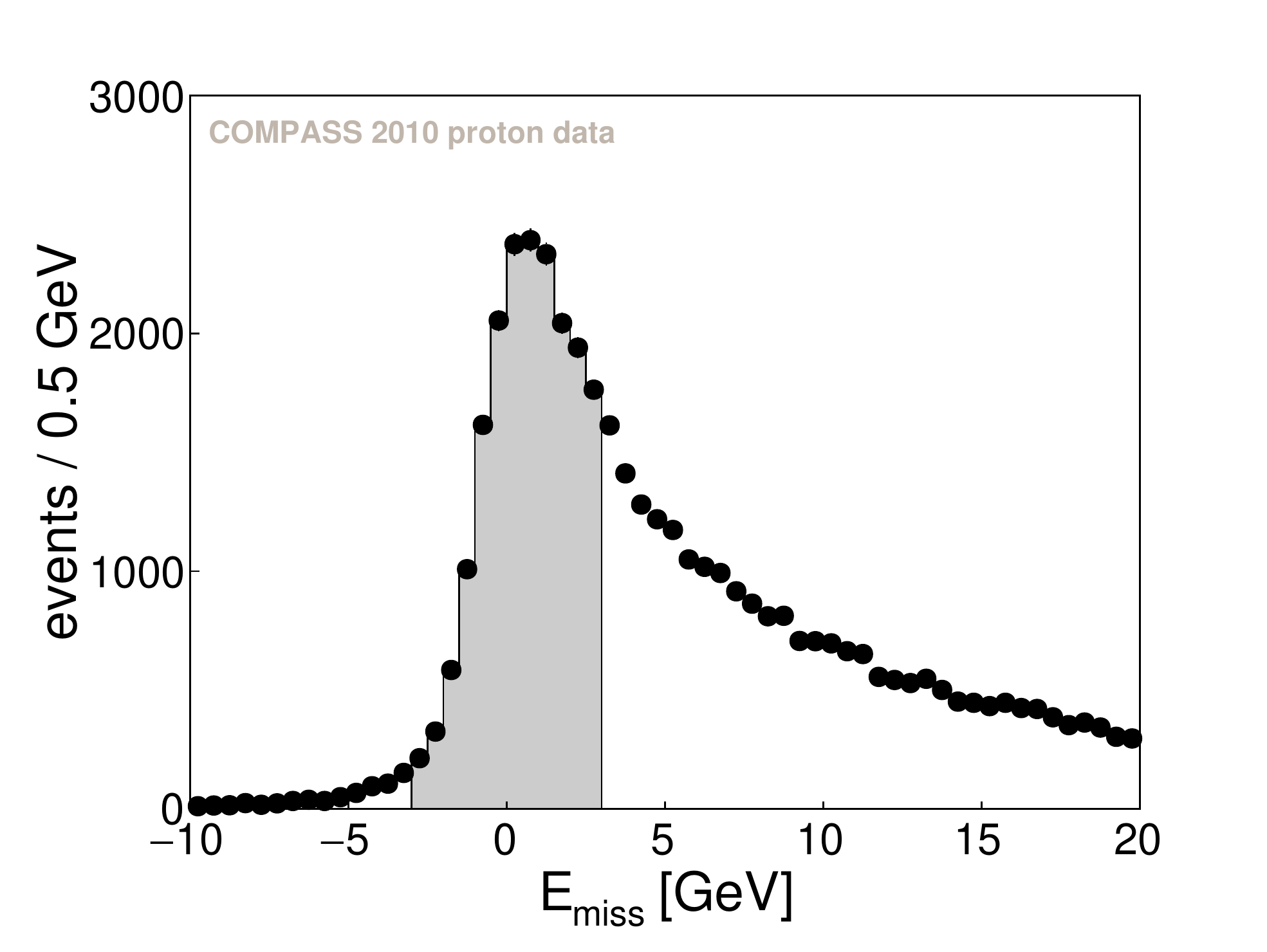}
\caption{Distribution of \varEmiss. The accepted events are denoted by the shaded area.}
\label{fig:selection:Emiss}
\end{figure}

As the recoiling proton is not detected, exclusive events are selected by the
cut on missing energy given by Eq.~\eqref{eq:selection:Emiss_limit}. The
selected range is referred to as `signal region' in the following. The
distribution of \varEmiss is shown in Fig.~\ref{fig:selection:Emiss}, where the
exclusive peak at $E_{\mathrm{miss}} \approx 0$ is clearly visible. The
boundaries of the \varEmiss range for the selection of exclusive events are
chosen to cover the $\pm 2\sigma$ region of the exclusive peak. Since it is not
possible to distinguish on an event-by-event basis between signal and
background events in the signal window, the background asymmetries are probed
in the second range of \varEmiss, 
\begin{equation}
7~\mathrm{GeV} < E_{\mathrm{miss}} < 20~\mathrm{GeV},
\label{eq:selection:Emiss_bkg_limit}
\end{equation}
where only semi-inclusive background events contribute. The intermediate range, $3~\mathrm{GeV} < E_{\mathrm{miss}} < 7~\mathrm{GeV}$, is contaminated by diffractive-dissociation events ($\gamma^{*}N\rightarrow\omega N^{*}$, where $N^{*}\rightarrow N+\pi+\hdots$), as indicated by results of Monte Carlo simulations \cite{Sandacz:2012at, Sjostrand:2006za}. Similarly as in the $\rho^0$ analysis \cite{Adolph:2013zaa}, this range is not taken into account in the present analysis. In order to reduce further the semi-inclusive background contribution, events are accepted only if the energy of the $\omega$ meson in the laboratory system is large enough, see Eq.~\eqref{eq:selection:Eomega_limit}.

Diffractive dissociation background in the exclusive sample is examined using a Monte Carlo event generator called HEPGEN \cite{Sandacz:2012at}. Using both exclusive and nucleon-dissociative $\omega$ events generated by HEPGEN, which are reconstructed and selected as the real data, the contribution from low-mass diffractive dissociation of the nucleon, corresponding to the \varEmiss range given in Eq.~\eqref{eq:selection:Emiss_limit}, is found to be $\approx 14\%$ of the exclusive $\omega$ signal.

\begin{figure}[tbp]
\centering
\includegraphics[width=.55\textwidth,trim=0 0 0 15]{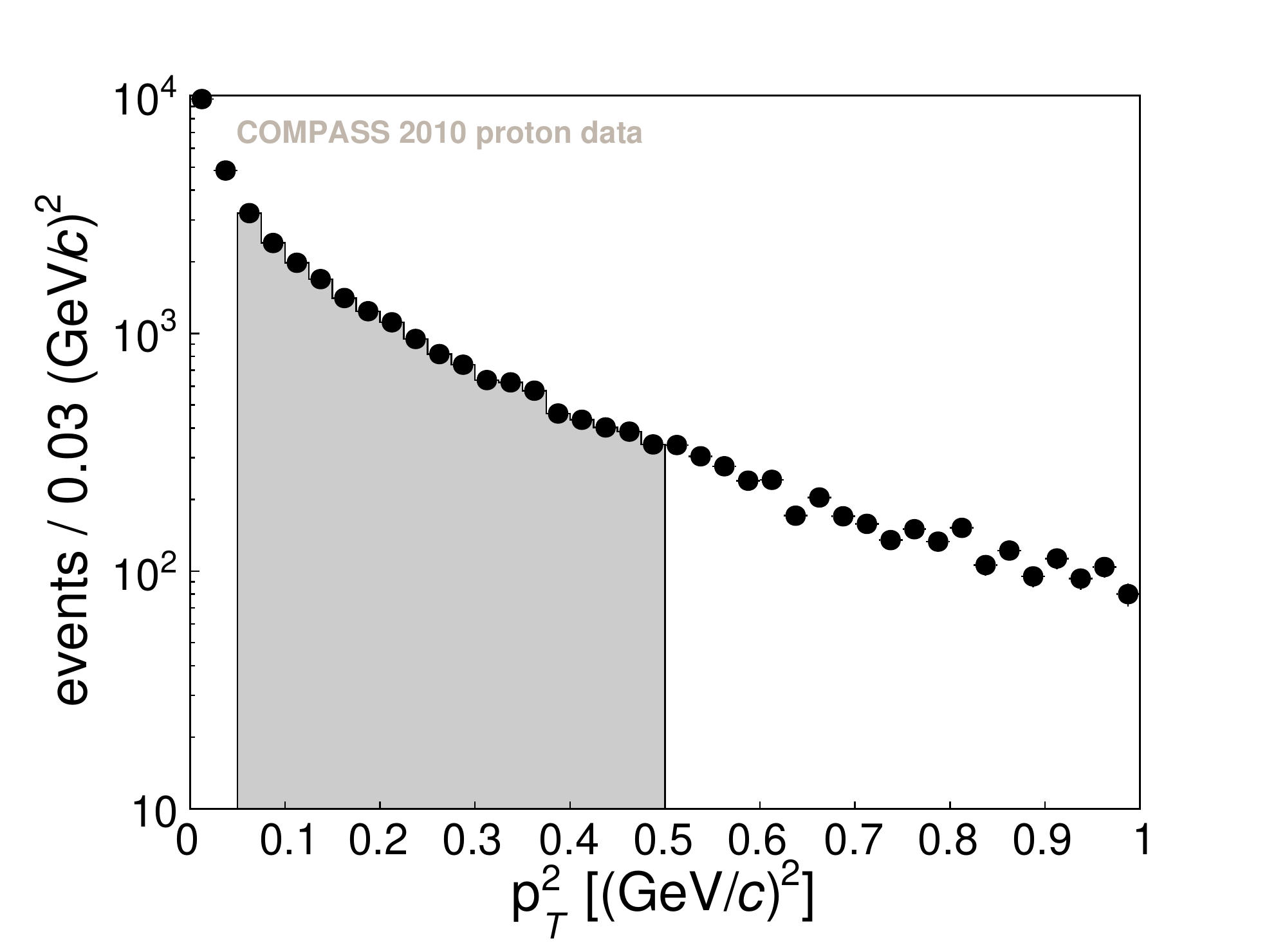}
\caption{Distribution of \varpttwo. 
The accepted events are denoted by the shaded area.}
\label{fig:selection:pt2}
\end{figure}

The \varpttwo distribution is shown in Fig.~\ref{fig:selection:pt2}. We choose
to use \varpttwo rather than \vart or $t'= |t| - t_{0}$, where $t_{0}$ is the
minimal kinematically allowed $|t|$. The reason is that in the COMPASS
kinematic region and for the set-up without detection of the recoil particle,
\varpttwo is determined with a precision better by a factor of two to five. In
addition, the $t'$ distribution is distorted because $t_{0}$, which depends on
\varW, \varQtwo, $M_{\pi^+\pi^-\pi^0}$ and $M_{X}^{2}$, is poorly determined
for non-exclusive background events \cite{Amaudruz:1991cc}. The \varpttwo
distribution shown in Fig.~\ref{fig:selection:pt2} indicates at small \varpttwo
values a contribution from coherent $\omega$ production on target nuclei.
Coherent events are suppressed by applying the lower limit given by
Eq.~\eqref{eq:selection:pt2_limit}. A study of \varpttwo distributions shows that
in addition to exclusive coherent and incoherent $\omega$ production a third
component, which originates from non-exclusive background, is also present and
its contribution increases with \varpttwo, thus requiring also an upper limit.
Therefore, in order to select the sample of events from incoherent exclusive
$\omega$ production, the afore mentioned \varpttwo limits are applied.

After all selections, the final sample for incoherent exclusive $\omega$
production consists of about $18500$ events. The mean values of the kinematic
variables \varQtwo, \varxbj, \vary, \varW and \varpttwo are given in 
Table~\ref{tab:selection:mean_kin_values}.

\begin{table}[t]
\centering
\caption{Mean values of selected kinematic variables for events reconstructed
in the signal region $-3~\mathrm{GeV} < E_{\mathrm{miss}} < 3~\mathrm{GeV}$
with and without the correction for semi-inclusive background
\cite{Sznajder:2015}.}
\label{tab:selection:mean_kin_values}
\begin{tabularx}{\textwidth}{X c c c c c}
\toprule
& $\langle Q^{2} \rangle$ $[(\mathrm{GeV}/\mathit{c})^2]$ & $\langle x_{\mathit{Bj}} \rangle$ & $\langle y\rangle$ & $\langle W \rangle$ $[\mathrm{GeV}/\mathit{c}^{2}]$ & $\langle p_{T}^{2} \rangle$ $[(\mathrm{GeV}/\mathit{c})^2]$\\
\midrule
signal only         & 2.2 & 0.049 & 0.18 & 7.1 & 0.17 \\
signal + background & 2.4 & 0.055 & 0.17 & 6.9 & 0.19 \\
\bottomrule
%&
%\multicolumn{1}{c}{\rule{0.09\textwidth}{0px}} &
%\multicolumn{1}{c}{\rule{0.09\textwidth}{0px}} &
%\multicolumn{1}{c}{\rule{0.09\textwidth}{0px}} &
%\multicolumn{1}{c}{\rule{0.09\textwidth}{0px}} &  
%\multicolumn{1}{c}{\rule{0.09\textwidth}{0px}} 
\end{tabularx}
\end{table}

\section{Extraction of asymmetries}
\label{sec:extraction}

The azimuthal asymmetries listed in Eq.~\eqref{eq:formalism:asymmetries} are
evaluated by fitting simultaneously the exclusive signal events (denoted by
subscript $\mathrm{S}$) and semi-inclusive background events (denoted by
subscript $\mathrm{B}$) using the unbinned maximum likelihood estimator. This
method of extraction allows us to study correlations between asymmetries and to
reduce the statistical uncertainty of the measurement compared to binned
estimators. 

Four subsamples of events are fitted simultaneously as a function of the
azimuthal angles and the missing energy. Each subsample corresponds to the
specific target cell $t$ with the polarisation state $p$. Here,
$t=\mathrm{C}$ and $t=\mathrm{U+D}$ refer to the central cell and the sum of
upstream and downstream cells, respectively, while the two target polarisation
states are denoted by $p=\uparrow$ and $p=\downarrow$. The fitted function
describes the observed sum of exclusive signal and semi-inclusive background
events denoted in the following by the subscript $\mathrm{S}+\mathrm{B}$: 
\begin{equation}
N_{t,\; \mathrm{S+B}}^{p}(\phi, \phi_{s}, E_{\mathrm{miss}}) =  
c_{t,\; \mathrm{S} + \mathrm{B}}^{p}(\phi, \phi_{s}, E_{\mathrm{miss}}) ~ 
g_{t,\; \mathrm{S+B}}^{p}(\phi, \phi_{s}, E_{\mathrm{miss}}). 
\label{eq:extraction:fit_function}
\end{equation}
In the factor
\begin{align}
c_{t,\; \mathrm{S} + \mathrm{B}}^{p}(\phi, \phi_{s}, E_{\mathrm{miss}}) = F ~ n ~ \left[ \sigma_{0,\; \mathrm{S}}^{\phantom{p}} ~ a_{t,\; \mathrm{S}}^{p}(\phi, \phi_{s}, E_{\mathrm{miss}}) + \sigma_{0,\; \mathrm{B}}^{\phantom{p}} ~ a_{t,\; \mathrm{B}}^{p}(\phi, \phi_{s}, E_{\mathrm{miss}}) \right] 
\label{eq:extraction:c_function}
\end{align}
$F$ is the muon flux, $n$ is the number of target nucleons, $\sigma_{0,\;
\mathrm{S}(\mathrm{B})}^{\phantom{p}}$ are the spin-averaged cross sections and
$a_{t,\; \mathrm{S}(\mathrm{B})}^{p}(\phi, \phi_{s}, E_{\mathrm{miss}})$ are
the acceptances for cell $t$ with polarisation $p$, where
$\mathrm{S}(\mathrm{B})$ denotes either $\mathrm{S}$ or $\mathrm{B}$. The
factor
\begin{align}
g_{t,\; \mathrm{S+B}}^{p}(\phi, \phi_{s}, E_{\mathrm{miss}}) = 1 \pm
\gamma_{t,\; \mathrm{S}}^{p}(\phi, \phi_{s}, E_{\mathrm{miss}}) ~
A_{\mathrm{raw},\; \mathrm{S}}(\phi, \phi_{s}) \pm \gamma_{t,\;
\mathrm{B}}^{p}(\phi, \phi_{s}, E_{\mathrm{miss}}) ~ A_{\mathrm{raw},\;
\mathrm{B}}(\phi,  \phi_{s}) 
\label{eq:extraction:g_function}
\end{align}
describes the measured azimuthal modulations of the cross section for
longitudinally polarised beam and transversely polarised target. In
Eq.~\eqref{eq:extraction:g_function}, 
\begin{equation}
\gamma_{t,\; \mathrm{S}(\mathrm{B})}^{p}(\phi, \phi_{s}, E_{\mathrm{miss}}) =
\frac{\sigma_{0,\; \mathrm{S}(\mathrm{B})}^{\phantom{p}} ~ a_{t,\;
\mathrm{S}(\mathrm{B})}^{p}(\phi, \phi_{s}, E_{\mathrm{miss}})}{\sigma_{0,\;
\mathrm{S}}^{\phantom{p}} ~ a_{t,\; \mathrm{S}}^{p}(\phi, \phi_{s},
E_{\mathrm{miss}}) + \sigma_{0,\; \mathrm{B}}^{\phantom{p}} ~ a_{t,\;
\mathrm{B}}^{p}(\phi, \phi_{s}, E_{\mathrm{miss}})} 
\end{equation}
are the weights corresponding to the fractions of signal and background
processes that are evaluated from the data as described in the following, while 
\begin{align}
A_{\mathrm{raw},\; \mathrm{S}(\mathrm{B})}(\phi, \phi_{s}) &=  
   A_{\mathrm{raw},\; \mathrm{S}(\mathrm{B})}^{\sin \left( \phi - \phi_{s} \right)} \sin \left( \phi - \phi_{s} \right) + 
   A_{\mathrm{raw},\; \mathrm{S}(\mathrm{B})}^{\sin \left( \phi + \phi_{s} \right)} \sin \left( \phi + \phi_{s} \right) + 
   A_{\mathrm{raw},\; \mathrm{S}(\mathrm{B})}^{\sin \left(2\phi - \phi_{s} \right)} \sin \left(2\phi - \phi_{s} \right) \nonumber \\
&+ A_{\mathrm{raw},\; \mathrm{S}(\mathrm{B})}^{\sin \left(3\phi - \phi_{s} \right)} \sin \left(3\phi - \phi_{s} \right) + 
   A_{\mathrm{raw},\; \mathrm{S}(\mathrm{B})}^{\sin \phi_{s}} \sin \phi_{s} +
   A_{\mathrm{raw},\; \mathrm{S}(\mathrm{B})}^{\cos \left( \phi - \phi_{s} \right)} \cos \left( \phi - \phi_{s} \right) \nonumber \\ 
&+ A_{\mathrm{raw},\; \mathrm{S}(\mathrm{B})}^{\cos \left(2\phi - \phi_{s} \right)} \cos \left(2\phi - \phi_{s} \right) +
   A_{\mathrm{raw},\; \mathrm{S}(\mathrm{B})}^{\cos \phi_{s}} \cos \phi_{s} 
\label{eq:extraction:all_asymmetries}
\end{align}
are the raw asymmetries that enter Eq.~\eqref{eq:extraction:g_function} with
the sign corresponding to the target polarisation state, $+$ and $-$ for
$\uparrow$ and $\downarrow$, respectively. The raw asymmetries are related to
the physics asymmetries, in particular to those defined in
Eq.~\eqref{eq:formalism:asymmetries} for the exclusive signal events, in the
following way:
\begin{align}
A_{\mathrm{UT},\; \mathrm{S}(\mathrm{B})}^{\mathrm{mod}} &= \frac{A_{\mathrm{raw},\; \mathrm{S}(\mathrm{B})}^{\mathrm{mod}}}{f_{\mathrm{S}(\mathrm{B})} ~ P_{T} ~ D^{\mathrm{mod}} }, \nonumber \\
A_{\mathrm{LT},\; \mathrm{S}(\mathrm{B})}^{\mathrm{mod}} &= \frac{A_{\mathrm{raw},\; \mathrm{S}(\mathrm{B})}^{\mathrm{mod}}}{f_{\mathrm{S}(\mathrm{B})} ~ P_{T} ~ P_{\ell} ~ D^{\mathrm{mod}}}. 
\label{eq:extraction:raw_asymmetries}
\end{align}
Here, the first line describes $\mathrm{UT}$ and the second one $\mathrm{LT}$
asymmetries, where `$\mathrm{mod}$' denotes the corresponding azimuthal
modulation and $f_{\mathrm{S}(\mathrm{B})}$ is the dilution factor defined in
Eq.~\eqref{eq:experiment:dilution_factor}. The target and beam polarisations
are given by $P_{T}$ and $P_{\ell}$, respectively. The depolarisation factors
$D^{\mathrm{mod}}$ depend on the virtual-photon polarisation parameter, see
Eq.~\eqref{eq:formalism:epsilon}: 
\begin{align}
D^{\sin \left( \phi - \phi_{S} \right)\text{\phantom{$3$}}} &= 1 ~, \nonumber \\
D^{\sin \left( \phi + \phi_{S} \right)\text{\phantom{$3$}}} &= D^{\sin \left(3\phi - \phi_{S} \right)} = \frac{\epsilon}{2} ~, \nonumber \\
D^{\sin \left(2\phi - \phi_{S} \right)} &= D^{\sin \phi_{S}} = \sqrt{\epsilon\left( 1 + \epsilon \right)} ~, \nonumber \\
D^{\cos \left( \phi - \phi_{S} \right)\text{\phantom{$3$}}} &= \sqrt{1-\epsilon^2} ~, \nonumber \\
D^{\cos \left(2\phi - \phi_{S} \right)} &= D^{\cos \phi_{S}} = \sqrt{\epsilon\left( 1 - \epsilon \right)}. 
\end{align}
In the fit of the function given by Eq.~\eqref{eq:extraction:fit_function}, the
unknowns are four functions $c_{t,\; \mathrm{S} + \mathrm{B}}^{p}(\phi,
\phi_{s}, E_{\mathrm{miss}})$ and sixteen physics asymmetries encoded in
$g_{t,\; \mathrm{S+B}}^{p}(\phi, \phi_{s}, E_{\mathrm{miss}})$. The other
parameters, \emph{i.e.}\ $\gamma_{t,\; \mathrm{S}(\mathrm{B})}^{p}$,
$f_{\mathrm{S}(\mathrm{B})}$, $P_{\ell}$ and $D^{\mathrm{mod}}$, are calculated
for each event, while $P_{T}$ is known from the target polarisation
measurement.

Equations~\eqref{eq:extraction:fit_function} to
\eqref{eq:extraction:all_asymmetries} are based upon two approximations:
\rNumIt{1}) the background asymmetries do not depend on the missing energy and
\rNumIt{2}) the smearing of azimuthal angles is neglected. Approximation
\rNumIt{1}) is justified by results of a study that revealed no dependence on
\varEmiss for asymmetries in the range $7~\mathrm{GeV} < E_{\mathrm{miss}} <
20~\mathrm{GeV}$, where only background events contribute. This observation
agrees with our previous analyses of exclusive $\rho^0$ production
\cite{Adolph:2012ht, Adolph:2013zaa}, where an analogous test was performed
with a much better statistical precision. A possible bias on the extraction of
asymmetries related to approximation \rNumIt{2}) is estimated in
Sec.~\ref{sec:systematics}. 

When fitting Eq.~\eqref{eq:extraction:fit_function}, one has to separate the
functions $c_{t,\; \mathrm{S} + \mathrm{B}}^{p}(\phi, \phi_{s},
E_{\mathrm{miss}})$ and $g_{t,\; \mathrm{S+B}}^{p}(\phi, \phi_{s},
E_{\mathrm{miss}})$ as due to the unknown acceptance both functions may be
correlated. The separation between both functions is achieved by using the
reasonable assumption that the ratio of acceptances in the cells stays the same
before and after target polarisation reversal:
\begin{equation} 
\frac{a_{\mathrm{U}+\mathrm{D},\; \mathrm{S}(\mathrm{B})}^{\uparrow}(\phi, \phi_{s}, E_{\mathrm{miss}})}{a_{\mathrm{C},\; \mathrm{S}(\mathrm{B})}^{\downarrow}(\phi, \phi_{s}, E_{\mathrm{miss}})} = 
\frac{a_{\mathrm{U}+\mathrm{D},\; \mathrm{S}(\mathrm{B})}^{\downarrow}(\phi, \phi_{s}, E_{\mathrm{miss}})}{a_{\mathrm{C},\; \mathrm{S}(\mathrm{B})}^{\uparrow}(\phi, \phi_{s}, E_{\mathrm{miss}})}.
\label{eq:extraction:reasonabe_assumption}
\end{equation}
If this assumption does not hold, false asymmetries may appear. Such a
possibility is examined in Sec.~\ref{sec:systematics}. 

Using different assumed functional forms of $c_{t,\; \mathrm{S} + \mathrm{B}}^{p}(\phi,
\phi_{s}, E_{\mathrm{miss}})$ in the fit has no significant effect on the
fitted parameters of the function $g_{t,\; \mathrm{S+B}}^{p}(\phi, \phi_{s},
E_{\mathrm{miss}})$, \emph{i.e.}\ on the physics asymmetries. Therefore, a
constant term is used in this analysis for the simplicity.

The possible dependence of $\gamma_{t,\; \mathrm{S}}^{p}(\phi, \phi_{s},
E_{\mathrm{miss}})$ and $\gamma_{t,\; \mathrm{B}}^{p}(\phi, \phi_{s},
E_{\mathrm{miss}})$ on the azimuthal angles is examined by using a Monte Carlo
(MC) simulation of the COMPASS apparatus. In this simulation the signal and
background processes were generated by HEPGEN \cite{Sandacz:2012at} and LEPTO
\cite{Ingelman:1996mq} generators, respectively. For the latter one the COMPASS
tuning \cite{Adolph:2012vj} of the JETSET parameters was used. The weights
$\gamma_{t,\; \mathrm{S}}^{p}(\phi, \phi_{s}, E_{\mathrm{miss}})$ and
$\gamma_{t,\; \mathrm{B}}^{p}(\phi, \phi_{s}, E_{\mathrm{miss}})$ are found to
be independent on the azimuthal angles.

\begin{figure}[tbp]
\centering
\includegraphics[width=0.55\textwidth,trim=0 0 0 15]{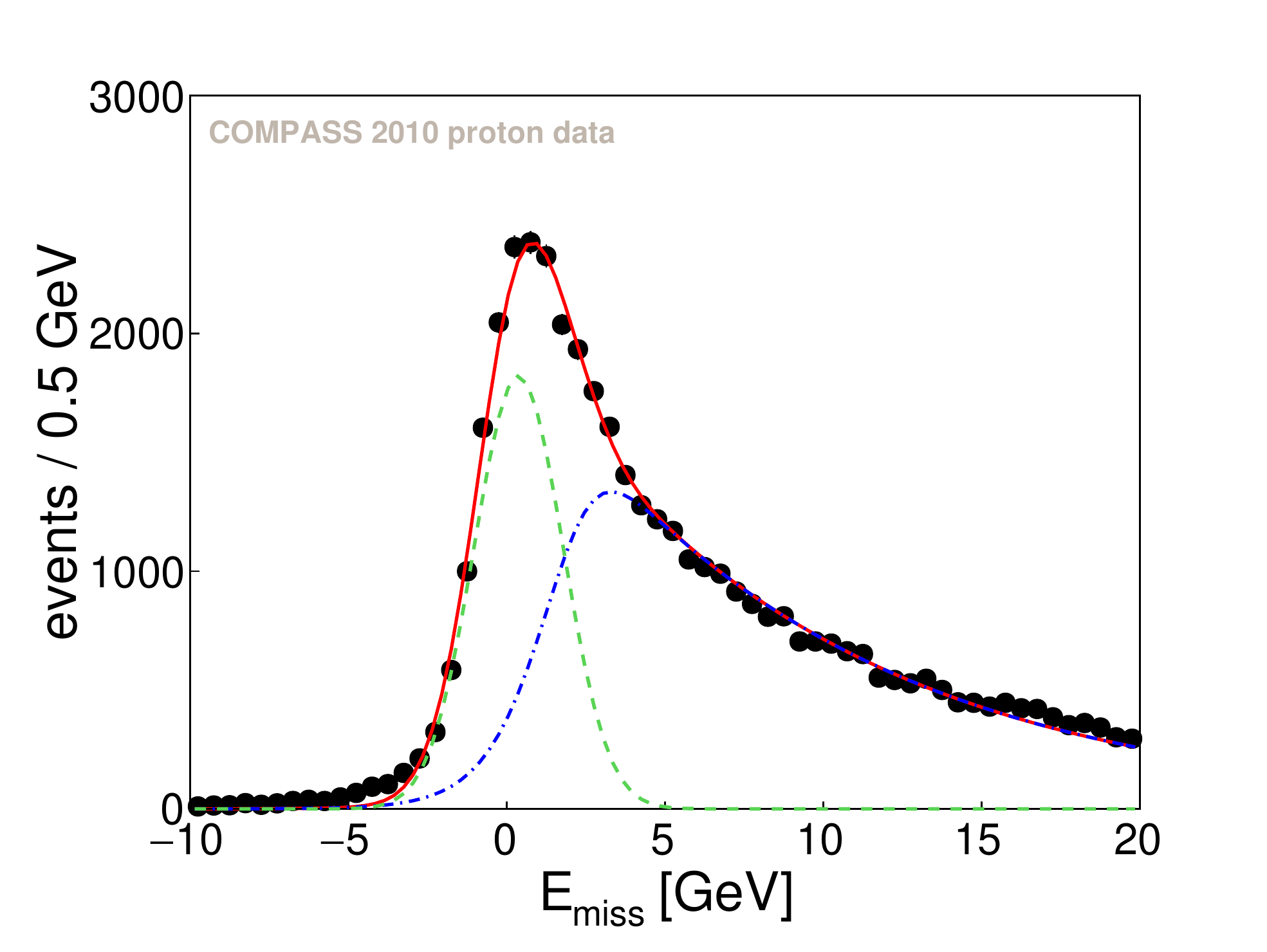}
\caption{Parameterised distribution of \varEmiss for the whole data sample. The
dashed (green) and dash-dotted (blue) curves represent the signal and
background contributions, respectively. The sum of both contributions is
represented by the solid (red) curve.} 
\label{fig:extraction:emiss_fits}
\end{figure}

The weights $\gamma_{t,\; \mathrm{S}}^{p}(E_{\mathrm{miss}})$ and $\gamma_{t,\;
\mathrm{B}}^{p}(E_{\mathrm{miss}}) = 1 - \gamma_{t,\;
\mathrm{S}}^{p}(E_{\mathrm{miss}})$ are calculated by parameterising the
missing energy distribution obtained for each target cell and each target
polarisation state, as illustrated in Fig.~\ref{fig:extraction:emiss_fits}. In
these parameterisations a Gaussian function is used for the shape of the
distribution of signal events, while for background events the shape is fixed
by the aforementioned MC simulation with LEPTO. In analogy to our previous
analyses \cite{Adolph:2012ht, Adolph:2013zaa}, the agreement between data and
MC events is improved by weighting each \varEmiss bin $i$ of the MC
distribution by the ratio 
\begin{equation}
w_{i} = \frac{
   N_{i,\; \mathrm{data}}^{\pi^{+}\pi^{+}\pi^{0}} + N_{i,\; \mathrm{data}}^{\pi^{-}\pi^{-}\pi^{0}}
}{
   N_{i,\; \mathrm{MC}}^{\pi^{+}\pi^{+}\pi^{0}} + N_{i,\; \mathrm{MC}}^{\pi^{-}\pi^{-}\pi^{0}}
}.
\end{equation}
Here, $N_{i,\; \mathrm{data}}^{\pi^{\pm}\pi^{\pm}\pi^{0}}$ and $N_{i,\;
\mathrm{MC}}^{\pi^{\pm}\pi^{\pm}\pi^{0}}$ are the numbers of events observed in
bin $i$ for experimental data and MC, respectively, when two hadrons with the
same charge are required in the selection of events. Such selection excludes
any exclusive production, so that the weights for semi-inclusive events can be
calculated at any value of \varEmiss.

\section{Systematic studies}
\label{sec:systematics}

In order to estimate the systematic uncertainties of this measurement the
following contributions were examined: \rNumIt{1}) false asymmetries,
\rNumIt{2}) a possible bias of the applied estimator of the asymmetries,
\rNumIt{3}) the sensitivity to the background parameterisation, \rNumIt{4}) the
stability of asymmetries over data taking time, \rNumIt{5}) the compatibility
between the three mean asymmetries obtained by averaging the one-dimensional
distributions in \varQtwo, \varxbj and \varpttwo, \rNumIt{6}) the uncertainty
in the calculation of dilution factor, beam and target polarisations.
\\ \\ 
\rNumIt{1}) 
False asymmetries are extracted by analysing subsamples of data with the same
spin orientation of target protons. In such a case, non-zero values of
azimuthal asymmetries would indicate an experimental bias. In particular, false
asymmetries provide a test of validity of the reasonable assumption, see
Eq.~\eqref{eq:extraction:reasonabe_assumption}, \emph{i.e.}\ whether during data
taking the acceptance has changed in a way that influences the extraction of
asymmetries. False asymmetries are determined in two ways: by using the data
from the upstream and downstream cells, as well as by the artificial division
of the central cell into two $30~\mathrm{cm}$ subcells. This test is performed
without separation into signal and background asymmetries, and independently
for the ranges $|E_{\mathrm{miss}}| < 3~\mathrm{GeV}$ and $-3~\mathrm{GeV} <
E_{\mathrm{miss}} < 20~\mathrm{GeV}$. The resulting false asymmetries are found
to be consistent with zero within statistical uncertainties. Nevertheless, an
upper limit on the false asymmetries is estimated to be $0.01$
\cite{Wolbeek:2015} at the level of raw asymmetries defined in
Eq.~\eqref{eq:extraction:raw_asymmetries}. This estimate represents a conservative
limit for breaking the reasonable assumption given by
Eq.~\eqref{eq:extraction:reasonabe_assumption}. At the level of physics
asymmetries, this estimation yields typically a systematic uncertainty on the
level of $20\%$ of the statistical one.  
\\ \\
\rNumIt{2})
The check of the extraction method is twofold. First, the effect of smearing
and acceptance on the extraction of asymmetries is examined by introducing an
asymmetry of known value to the MC data generated with the HEPGEN generator
\cite{Sandacz:2012at}, which is followed by the simulation of the COMPASS
apparatus and event reconstruction. It is checked whether the unbinned maximum
likelihood estimator returns the introduced asymmetry correctly. In addition, a
possible mixing between asymmetries is investigated, \emph{i.e.}\ it is checked
whether a non-zero asymmetry contributes to any other asymmetry. The test shows
only an effect of smearing. The related systematic uncertainty is estimated to
be up to $33\%$ of the statistical one. 

In the second test the obtained asymmetries are compared with those extracted
with an alternative estimator that is chosen to be the 2D binned maximum
likelihood estimator. In this case, in contrast to that of the unbinned
estimator, the extraction of asymmetries proceeds after performing the
subtraction of semi-inclusive background that is probed in the range
$7~\mathrm{GeV} < E_{\mathrm{miss}} < 20~\mathrm{GeV}$. This way of extraction
was used in the COMPASS analysis of transverse target spin asymmetries for the
$\rho^0$ meson \cite{Adolph:2013zaa}. The comparison indicates a good agreement
between both estimators.  
\\ \\
\rNumIt{3})
The systematic uncertainty related to the background treatment, see
Sec.~\ref{sec:extraction}, is estimated by assuming that the fractions of signal and
background processes are known with an uncertainty of $10 \%$. This assumption
is supported by the COMPASS analysis for the $\rho^0$ meson
\cite{Adolph:2013zaa}, where the MC samples produced with LEPTO
\cite{Ingelman:1996mq} and PYTHIA \cite{Sjostrand:2006za} generators were
compared. The estimation yields typically a systematic uncertainty of $30\%$ of
the statistical one.   
\\ \\
\rNumIt{4})
The data stability over time is examined by comparing asymmetries extracted
from two consecutive subsets of data taking. A division into a larger number of
subsets is not possible due to limited statistics. All asymmetries extracted
from the two subsets of data are found to be compatible within statistical
uncertainties.
\\ \\
\rNumIt{5})
The extraction of asymmetries may be unstable due to limited statistics. The
effect is examined by comparing the asymmetries extracted from the entire data
sample with those obtained from averaging the results obtained in bins of
kinematic variables \varQtwo, \varxbj or \varpttwo, which are used for the
extraction of final results, see Fig.~\ref{fig:results:results} (right). The
comparison indicates a bias that is estimated to be up to $20\%$ of the
statistical uncertainty.  
\\ \\
\rNumIt{6})
In order to estimate the normalisation (scale) systematic uncertainty, we take
into account the relative uncertainty of the target dilution factor, $2\%$, the
target polarisation, $3\%$, and the beam polarisation, $5\%$. Combined in
quadrature, this yields an overall systematic normalisation uncertainty of
$3.6\%$ for the single-spin asymmetries and $6.2\%$ for the double-spin ones. 

The systematic uncertainties for the average asymmetries obtained in
\rNumIt{1}) - \rNumIt{5}) are summarised in
Table~\ref{tab:systematics:summary}. The total systematic uncertainties evaluated by
summing in quadrature the values obtained in \rNumIt{1}) - \rNumIt{6}) are
given in Table~\ref{tab:results:summary}.

\begin{table}[t]
\centering
\caption{Systematic uncertainties for the average asymmetries obtained from the
studies explained in the text. The uncertainties related to \rNumIt{4}) are
negligible for all asymmetries. The scaling uncertainties are not included in
this table.}
\label{tab:systematics:summary}
\begin{tabularx}{\textwidth}{l c c c c X l c c c c}
\toprule
   & \rNumIt{1}) & \rNumIt{2}) & \rNumIt{3}) & \rNumIt{5}) & & & \rNumIt{1}) & \rNumIt{2}) & \rNumIt{3}) & \rNumIt{5}) \\
\midrule
$A_{\mathrm{UT}}^{\sin \left( \phi - \phi_{S} \right)}$ & 0.015& 0.002& 0.025& 0.012& & 
$A_{\mathrm{LT}}^{\cos \left( \phi - \phi_{S} \right)}$ & 0.077& 0.001& 0.126& 0.109\\
$A_{\mathrm{UT}}^{\sin \left( \phi + \phi_{S} \right)}$ & 0.029& 0.001& 0.048& 0.021& &
$A_{\mathrm{LT}}^{\cos \left(2\phi - \phi_{S} \right)}$ & 0.114& 0.002& 0.181& 0.108\\
$A_{\mathrm{UT}}^{\sin \left(2\phi - \phi_{S} \right)}$ & 0.011& 0.010& 0.027& 0.006& &
$A_{\mathrm{LT}}^{\cos \phi_{S}}$                       & 0.115& 0.076& 0.179& 0.123\\
$A_{\mathrm{UT}}^{\sin \left(3\phi - \phi_{S} \right)}$ & 0.029& 0.049& 0.051& 0.003& &
& & & & \\
$A_{\mathrm{UT}}^{\sin \phi_{S}}$                       & 0.010& 0.015& 0.019& 0.010&
& & & & \\
\bottomrule
%\multicolumn{1}{c}{\rule{0.12\textwidth}{0px}} &
%\multicolumn{1}{c}{\rule{0.05\textwidth}{0px}} &
%\multicolumn{1}{c}{\rule{0.05\textwidth}{0px}} &
%\multicolumn{1}{c}{\rule{0.05\textwidth}{0px}} &
%\multicolumn{1}{c}{\rule{0.05\textwidth}{0px}} & & 
%\multicolumn{1}{c}{\rule{0.12\textwidth}{0px}} &
%\multicolumn{1}{c}{\rule{0.05\textwidth}{0px}} &
%\multicolumn{1}{c}{\rule{0.05\textwidth}{0px}} &
%\multicolumn{1}{c}{\rule{0.05\textwidth}{0px}} &
%\multicolumn{1}{c}{\rule{0.05\textwidth}{0px}} 
\end{tabularx}
\end{table}

\begin{table}[t]
\centering
\caption{Average azimuthal asymmetries for exclusive $\omega$ muoproduction with statistical and systematic uncertainties for all measured modulations.}
\label{tab:results:summary}
\begin{tabularx}{\textwidth}{l l l l X l l l l}
\toprule
   & \phantom{$-$}$\mathrm{A}$ & $\sigma_{\mathrm{stat}}$ & $\sigma_{\mathrm{sys}}$  & & & $\mathrm{A}$ & $\sigma_{\mathrm{stat}}$ & $\sigma_{\mathrm{sys}}$ \\
\midrule
$A_{\mathrm{UT}}^{\sin \left( \phi - \phi_{S} \right)}$ &$-0.059$& 0.074& 0.031& &
$A_{\mathrm{LT}}^{\cos \left( \phi - \phi_{S} \right)}$ &$0.07$  & 0.42& 0.18\\
$A_{\mathrm{UT}}^{\sin \left( \phi + \phi_{S} \right)}$ &\phantom{$-$}$0.06$  & 0.15& 0.06& &
$A_{\mathrm{LT}}^{\cos \left(2\phi - \phi_{S} \right)}$ &$0.01$  & 0.61& 0.24\\
$A_{\mathrm{UT}}^{\sin \left(2\phi - \phi_{S} \right)}$ &$-0.054$& 0.053& 0.031& &
$A_{\mathrm{LT}}^{\cos \phi_{S}}$                       &$0.54  $& 0.58& 0.26\\
$A_{\mathrm{UT}}^{\sin \left(3\phi - \phi_{S} \right)}$ &\phantom{$-$}$0.13  $& 0.15& 0.08& &
& & & \\
$A_{\mathrm{UT}}^{\sin \phi_{S}}$                       & \phantom{$-$}0.096& 0.059& 0.028&
& & & \\
\bottomrule
%\multicolumn{1}{c}{\rule{0.12\textwidth}{0px}} &
%\multicolumn{1}{c}{\rule{0.08\textwidth}{0px}} &
%\multicolumn{1}{c}{\rule{0.08\textwidth}{0px}} &
%\multicolumn{1}{c}{\rule{0.08\textwidth}{0px}} & & 
%\multicolumn{1}{c}{\rule{0.12\textwidth}{0px}} &
%\multicolumn{1}{c}{\rule{0.08\textwidth}{0px}} &
%\multicolumn{1}{c}{\rule{0.08\textwidth}{0px}} &
%\multicolumn{1}{c}{\rule{0.08\textwidth}{0px}} 
\end{tabularx}
\end{table}

\section{Results and discussion}
\label{sec:results}

\begin{figure}[tbp]
\centering 
\raisebox{-0.5\height}{\includegraphics[width=0.45\textwidth] {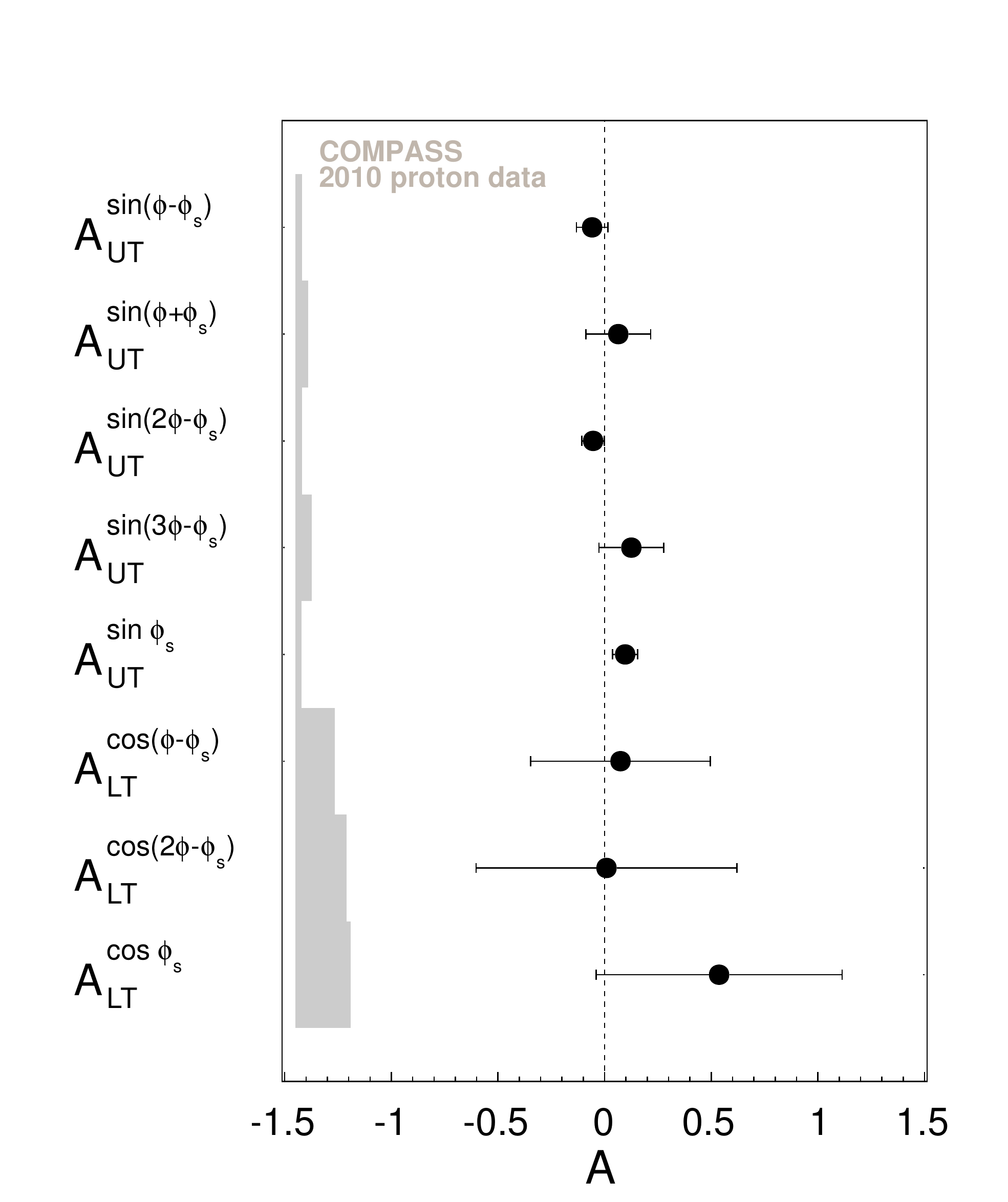}}
\raisebox{-0.5\height}{\includegraphics[width=0.45\textwidth] {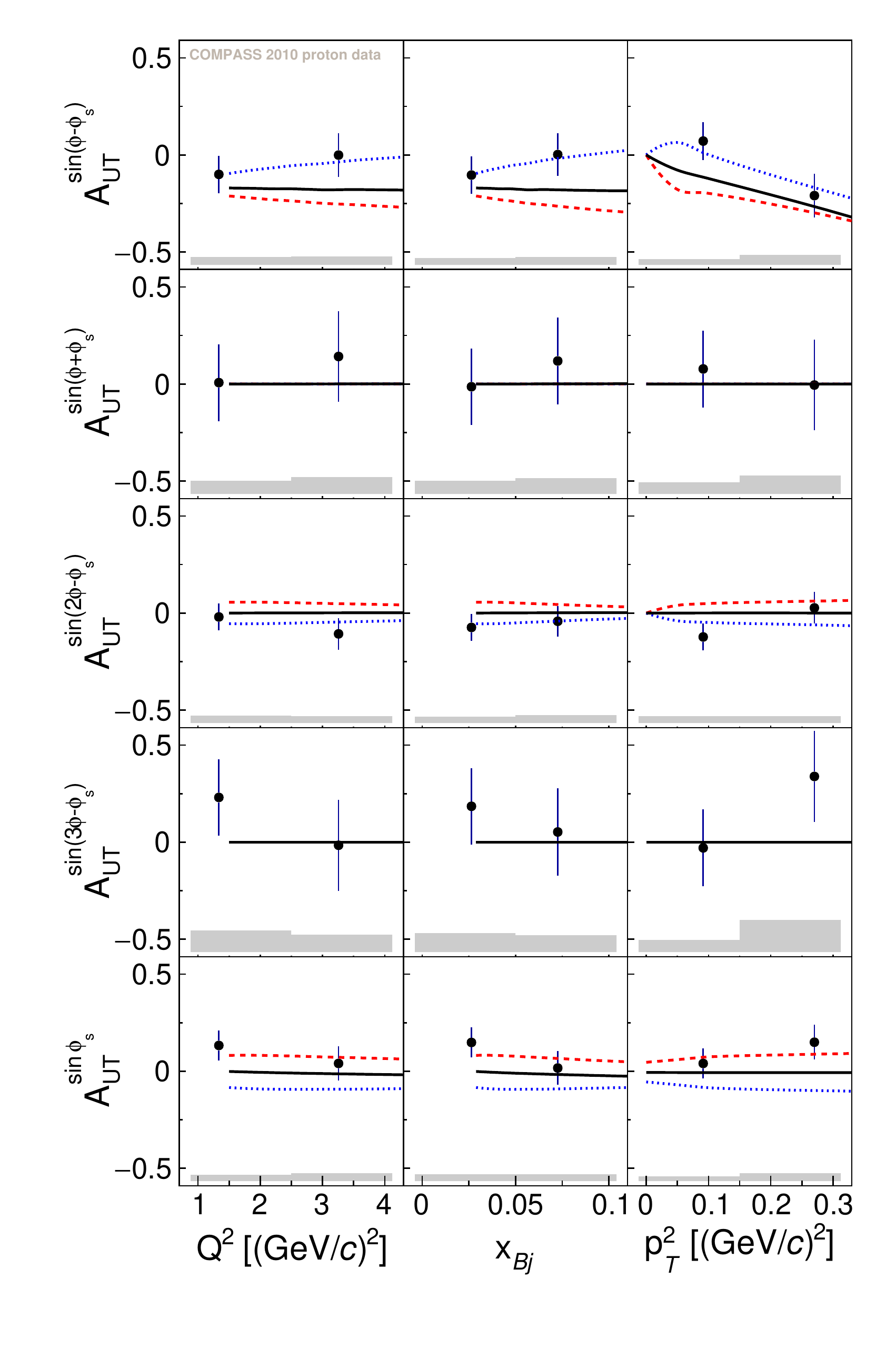}}
\caption{Left: Average azimuthal asymmetries for exclusive $\omega$
muoproduction. The error bars (left bands) represent the statistical
(systematic) uncertainties. Right: Single spin azimuthal asymmetries as a
function of \varQtwo, \varxbj and \varpttwo. The curves show the predictions of
the GPD-based model \cite{Goloskokov:2014ika} for the average \varQtwo, \varW
and \varpttwo values of the COMPASS data. The dashed red and dotted blue curves
represent the predictions with the positive and negative $\pi\omega$ form
factors, respectively, while the solid black curve represents the predictions
without the pion pole.} 
\label{fig:results:results}
\end{figure}

The measured azimuthal asymmetries, averaged over the entire kinematic range,
are given in Table~\ref{tab:results:summary} and shown in
Fig.~\ref{fig:results:results} (left). In addition, the single-spin asymmetries are
measured in bins of \varQtwo, \varxbj or \varpttwo with the results shown in
Fig.~\ref{fig:results:results} (right). The double-spin asymmetries are not
presented in separate kinematic bins because of large uncertainties. All
published results are available in the Durham data base \cite{Durham:db}.

In Figure~\ref{fig:results:results} (right) the COMPASS results are compared to
the calculations of the GK model \cite{Goloskokov:2014ika}. The latter are
obtained for the average \varW, \varQtwo and \varpttwo values of the COMPASS
data: $W = 7.1~\mathrm{GeV}/\mathit{c}^{2}$ and $p_{T}^{2} =
0.17~(\mathrm{GeV}/\mathit{c})^{2}$ for the \varxbj and \varQtwo dependences,
and $W = 7.1~\mathrm{GeV}/\mathit{c}^{2}$ and $Q^{2} =
2.2~(\mathrm{GeV}/c)^{2}$ for the \varpttwo dependence. The predictions are
given for three versions of the model: with the pion-pole contribution using a
positive or negative $\pi\omega$ transition form factor, and without the
pion-pole contribution.

\begin{figure}[tbp]
\centering 
\includegraphics[width=0.45\textwidth] {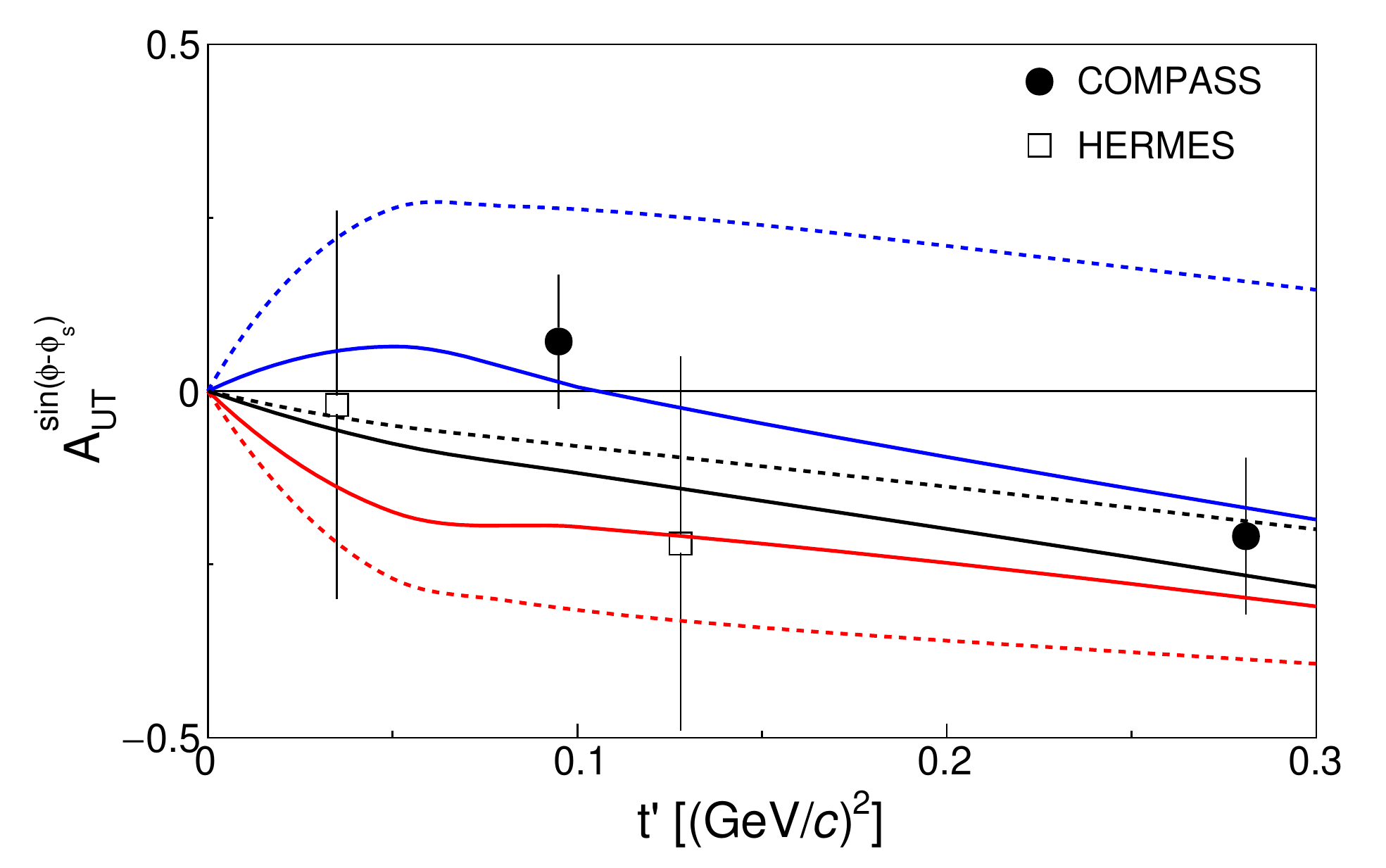}
\caption{The asymmetry \assAsfmfs for exclusive $\omega$ muoproduction measured
by the COMPASS (filled circles) and HERMES \cite{Airapetian:2015jxa} (open
squares) collaborations as a function of $t'$. The curves show the predictions
of the GPD-based model \cite{Goloskokov:2014ika} given for the average \varQtwo
and \varW values of the COMPASS (solid lines) and HERMES (dashed lines) data.
For each set of curves, the upper (blue) and lower (red) ones are for the
negative and positive $\pi\omega$ form factors, respectively, while the middle
(black) one represents the predictions without the pion pole.} 
\label{fig:results:HERMESvsCOMPASS}
\end{figure}

The asymmetry \assAsfmfs for exclusive $\omega$ production predicted by the
model without pion-pole contribution is $-0.11$. This value is significantly
different from that for exclusive $\rho^0$ production, which amounts to
$-0.01$. Thus in principle a combined analysis of results for this asymmetry
for both mesons could allow for a separation of the contributions of GPDs $E^{u}$
and $E^{d}$, which are different in both cases, as mentioned in the
Introduction. 

However, the interpretation of $\omega$ results in the context of the GPD
formalism is more challenging than that for $\rho^0$, as exclusive $\omega$
meson production is significantly influenced by the pion-pole exchange
contribution, and at present the sign of $\pi\omega$ transition form factor is
unknown. By comparing the COMPASS results with the calculations of the GK model
(see Fig.~\ref{fig:results:results} (right)), one finds that the asymmetries
\assAsfmfs and \assAsdfmfs prefer the negative $\pi\omega$ transition form
factor, while the asymmetry \assAsfs prefers the positive one. The other
measured asymmetries are not sensitive to the sign of the $\pi\omega$ form
factor. 

The single-spin azimuthal asymmetries for $\omega$ production on transversely
polarised protons were measured also by the HERMES collaboration
\cite{Airapetian:2015jxa}. They conclude that these data seem to favour the
positive $\pi\omega$ form factor, although within large experimental
uncertainties. A direct comparison of published asymmetry values measured in
both experiments in not straightforward, because the HERMES definition of
physics asymmetries differs from that given in
Eq.~\eqref{eq:formalism:asymmetries}. Such comparison is only possible for the
asymmetry \assAsfmfs. The results from both experiments are shown as a function
of $t'$ in Fig.~\ref{fig:results:HERMESvsCOMPASS} indicating their
compatibility within experimental uncertainties. Note that the COMPASS results
cover a wider kinematic range and they have smaller uncertainties, for example
for the asymmetry \assAsfmfs by a factor larger than two. 

The next measurement of exclusive meson production on a transversely polarised
target is expected to be performed at Jefferson Lab after the $12~\mathrm{GeV}$
upgrade \cite{Dudek:2012vr}. The foreseen data, although to be taken at
different kinematics, may contribute to the determination of the sign of the
$\pi\omega$ transition form factor. There are also plans to measure hard
exclusive meson production on transversely polarised protons by combining a
transversely polarised target with a recoil proton detector using an upgraded
COMPASS set-up \cite{Gautheron:2010wva}.

\section{Acknowledgements}
\label{sec:acknowledgements}

We gratefully acknowledge the support of the CERN management and staff and the
skill and effort of the technicians of our collaborating institutes. This work
was made possible by the financial support of our funding agencies. Special
thanks go to S. V. Goloskokov and P. Kroll for providing us with the full set
of model calculations as well as for the fruitful collaboration and many
discussions on the interpretation of the results.
\clearpage

%\bibliography{bibliography/bibliography}
\bibliography{bibliography/bib2}

\begin{thebibliography}{36}
\providecommand{\natexlab}[1]{#1}
\providecommand{\url}[1]{\texttt{#1}}
\expandafter\ifx\csname urlstyle\endcsname\relax
  \providecommand{\doi}[1]{doi: #1}\else
  \providecommand{\doi}{doi: \begingroup \urlstyle{rm}\Url}\fi

\bibitem{Mueller:1998fv}
D.~Mueller, D.~Robaschik, B.~Geyer, F.~M. Dittes, and J.~Horejsi, {Wave
  functions, evolution equations and evolution kernels from light ray operators
  of QCD}, \emph{Fortsch. Phys.} {\bf 42}, \penalty0 101--141,  (1994).

\bibitem{Ji:1996ek}
X.-D. Ji, {Gauge-invariant decomposition of nucleon spin}, \emph{Phys. Rev.
  Lett.} {\bf 78}, \penalty0 610--613,  (1997).

\bibitem{Ji:1996nm}
X.-D. Ji, {Deeply virtual Compton scattering}, \emph{Phys. Rev.} {\bf D55},
  \penalty0 7114--7125,  (1997).

\bibitem{Radyushkin:1996ru}
A.~V. Radyushkin, {Asymmetric gluon distributions and hard diffractive
  electroproduction}, \emph{Phys. Lett.} {\bf B385}, \penalty0 333--342,
  (1996).

\bibitem{Radyushkin:1997ki}
A.~V. Radyushkin, {Nonforward parton distributions}, \emph{Phys. Rev.} {\bf
  D56}, \penalty0 5524--5557,  (1997).

\bibitem{Burkardt:2000za}
M.~Burkardt, {Impact parameter dependent parton distributions and off forward
  parton distributions for $\zeta \rightarrow 0$}, \emph{Phys. Rev.} {\bf D62},
  \penalty0 071503,  (2000).
\newblock [Erratum-ibid.: {\bf{D66}}, 119903 (2002)].

\bibitem{Burkardt:2002hr}
M.~Burkardt, {Impact parameter space interpretation for generalized parton
  distributions}, \emph{Int. J. Mod. Phys.} {\bf A18}, \penalty0 173--208,
  (2003).

\bibitem{Burkardt:2004bv}
M.~Burkardt, {Generalized parton distributions for large $x$}, \emph{Phys.
  Lett.} {\bf B595}, \penalty0 245--249,  (2004).

\bibitem{Collins:1996fb}
J.~C. Collins, L.~Frankfurt, and M.~Strikman, {Factorization for hard exclusive
  electroproduction of mesons in QCD}, \emph{Phys. Rev.} {\bf D56}, \penalty0
  2982--3006,  (1997).

\bibitem{Martin:1996bp}
A.~D. Martin, M.~G. Ryskin, and T.~Teubner, {The QCD description of diffractive
  $\rho$ meson electroproduction}, \emph{Phys. Rev.} {\bf D55}, \penalty0
  4329--4337,  (1997).

\bibitem{Goloskokov:2005sd}
S.~V. Goloskokov and P.~Kroll, {Vector meson electroproduction at small
  Bjorken-$x$ and generalized parton distributions}, \emph{Eur. Phys. J.} {\bf
  C42}, \penalty0 281--301,  (2005).

\bibitem{Goloskokov:2007nt}
S.~V. Goloskokov and P.~Kroll, {The role of the quark and gluon GPDs in hard
  vector-meson electroproduction}, \emph{Eur. Phys. J.} {\bf C53}, \penalty0
  367--384,  (2008).

\bibitem{Goloskokov:2008ib}
S.~V. Goloskokov and P.~Kroll, {The target asymmetry in hard vector-meson
  electroproduction and parton angular momenta}, \emph{Eur. Phys. J.} {\bf
  C59}, \penalty0 809--819,  (2009).

\bibitem{Goloskokov:2013mba}
S.~V. Goloskokov and P.~Kroll, {Transversity in exclusive vector-meson
  leptoproduction}, \emph{Eur. Phys. J.} {\bf C74}, \penalty0 2725,  (2014).
\newblock and private communication.

\bibitem{Goloskokov:2014ika}
S.~V. Goloskokov and P.~Kroll, {The pion pole in hard exclusive vector-meson
  leptoproduction}, \emph{Eur. Phys. J.} {\bf A50}\penalty0 (9), \penalty0 146,
   (2014).
\newblock and private communication.

\bibitem{Adolph:2012ht}
C.~Adolph et~al., {Exclusive $\rho^0$ muoproduction on transversely polarised
  protons and deuterons}, \emph{Nucl. Phys.} {\bf B865}, \penalty0 1--20,
  (2012).

\bibitem{Adolph:2013zaa}
C.~Adolph et~al., {Transverse target spin asymmetries in exclusive $\rho^0$
  muoproduction}, \emph{Phys. Lett.} {\bf B731}, \penalty0 19--26,  (2014).

\bibitem{Diehl:2003ny}
M.~Diehl, {Generalized parton distributions}, \emph{Phys. Rep.} {\bf 388},
  \penalty0 41--277,  (2003).

\bibitem{Diehl:2004wj}
M.~Diehl and A.~V. Vinnikov, {Quarks vs. gluons in exclusive $rho$
  electroproduction}, \emph{Phys. Lett.} {\bf B609}, \penalty0 286--290,
  (2005).

\bibitem{Goloskokov:2006hr}
S.~V. Goloskokov and P.~Kroll, {The Longitudinal cross-section of vector meson
  electroproduction}, \emph{Eur. Phys. J.} {\bf C50}, \penalty0 829--842,
  (2007).

\bibitem{Bauer:1977iq}
T.~H. Bauer, R.~D. Spital, D.~R. Yennie, and F.~M. Pipkin, {The hadronic
  properties of the photon in high-energy interactions}, \emph{Rev. Mod. Phys.}
  {\bf 50}, \penalty0 261,  (1978).
\newblock [Erratum-ibid.: {\bf{51}}, 407 (1979)].

\bibitem{Airapetian:2014gfp}
A.~Airapetian et~al., {Spin density matrix elements in exclusive $\omega$
  electroproduction on $^1$H and $^2$H targets at $27.5$ GeV beam energy},
  \emph{Eur. Phys. J.} {\bf C74}\penalty0 (11), \penalty0 3110,  (2014).

\bibitem{Diehl:2005pc}
M.~Diehl and S.~Sapeta, {On the analysis of lepton scattering on longitudinally
  or transversely polarized protons}, \emph{Eur. Phys. J.} {\bf C41}, \penalty0
  515--533,  (2005).

\bibitem{Abbon:2007pq}
P.~Abbon et~al., {The COMPASS experiment at CERN}, \emph{Nucl. Instrum. Meth.}
  {\bf A577}, \penalty0 455--518,  (2007).

\bibitem{Alexakhin:2007mw}
V.~{\relax Yu}. Alexakhin et~al., {Double spin asymmetry in exclusive
  $\rho^{0}$ muoproduction at COMPASS}, \emph{Eur. Phys. J.} {\bf C52},
  \penalty0 255--265,  (2007).

\bibitem{Wolbeek:2015}
J.~ter Wolbeek.
\newblock \emph{{Azimuthal asymmetries in hard exclusive meson muoproduction
  off transversely polarized protons}}.
\newblock PhD thesis, {Albert-Ludwigs-Universit{\"a}t, Freiburg},  (2015).

\bibitem{Sandacz:2012at}
A.~Sandacz and P.~Sznajder, {HEPGEN - generator for hard exclusive
  leptoproduction}.  (2012).

\bibitem{Sjostrand:2006za}
T.~Sjostrand, S.~Mrenna, and P.~Z. Skands, {PYTHIA 6.4 Physics and Manual},
  \emph{JHEP}. {\bf 05}, \penalty0 026,  (2006).

\bibitem{Amaudruz:1991cc}
P.~Amaudruz et~al., {Transverse momentum distributions for exclusive $\rho^{0}$
  muoproduction}, \emph{Z. Phys.} {\bf C54}, \penalty0 239--246,  (1992).

\bibitem{Sznajder:2015}
P.~Sznajder.
\newblock \emph{{Study of azimuthal asymmetries in exclusive leptoproduction of
  vector mesons on transversely polarised protons and deuterons}}.
\newblock PhD thesis, {National Centre for Nuclear Research, Warsaw},  (2015).

\bibitem{Ingelman:1996mq}
G.~Ingelman, A.~Edin, and J.~Rathsman, {LEPTO 6.5: A Monte Carlo generator for
  deep inelastic lepton - nucleon scattering}, \emph{Comput. Phys. Commun.}
  {\bf 101}, \penalty0 108--134,  (1997).

\bibitem{Adolph:2012vj}
C.~Adolph et~al., {Leading order determination of the gluon polarisation from
  DIS events with high-$p_T$ hadron pairs}, \emph{Phys. Lett.} {\bf B718},
  \penalty0 922--930,  (2013).

\bibitem{Durham:db}
{The Durham HepData Project}.
\newblock \url{http://hepdata.cedar.ac.uk}.

\bibitem{Airapetian:2015jxa}
A.~Airapetian et~al., {Transverse-target-spin asymmetry in exclusive
  $\omega$-meson electroproduction}, \emph{Eur. Phys. J.} {\bf C75}, \penalty0
  600,  (2015).

\bibitem{Dudek:2012vr}
J.~Dudek et~al., {Physics Opportunities with the 12 GeV Upgrade at Jefferson
  Lab}, \emph{Eur. Phys. J.} {\bf A48}, \penalty0 187,  (2012).

\bibitem{Gautheron:2010wva}
F.~Gautheron et~al.
\newblock {COMPASS-II Proposal}.
\newblock Technical Report CERN-SPSC-2010-014. SPSC-P-340, CERN, Geneva (May,
  2010).

\end{thebibliography}
\bibliographystyle{ws-rv-van}

\end{document}